\documentclass[12pt, draftclsnofoot, onecolumn]{IEEEtran}
\usepackage{soul}            % for \st (strikeout command)
\usepackage{multicol}
\usepackage{epsfig}
\usepackage{epstopdf}
\usepackage{float}
\usepackage{subfig}
\usepackage{perpage}
\MakeSorted{table}
\usepackage{array}
\usepackage{relsize}
\usepackage{tikz}
\usepackage[linesnumbered,ruled]{algorithm2e}
\usepackage{enumitem}
\usepackage{graphicx}
\usepackage{amsfonts}
\usepackage{amssymb}

\usepackage[T1]{fontenc}
\usepackage{etoolbox}
\usepackage[noend]{algpseudocode}
\usepackage{cite}
\DeclareMathAlphabet\mathbfcal{OMS}{cmsy}{b}{n}
\usepackage[cmex10]{amsmath}
\usepackage{array}
\usepackage{fixltx2e}
\usepackage{wrapfig}
\usepackage[hidelinks]{hyperref}
\usepackage{mathtools, cuted}
\usepackage{hhline}
    \usepackage{pifont}
    \newcommand{\tick}{\ding{52}}

\usepackage{array}% http://ctan.org/pkg/array

\ifCLASSINFOpdf
\else
\fi

\usepackage{etoolbox} % for \iftoggle
\newtoggle{SINGLE_COL}
\toggletrue{SINGLE_COL}     % for single column format
\togglefalse{SINGLE_COL}   % for two column format
\newtheorem{lemma}{Lemma}

\makeatletter
\def\BState{\State\hskip-\ALG@thistlm}
\makeatother

\newtheorem{observation}{Observation}
\newtheorem{remark}{Remark}

\begin{document}
\bstctlcite{IEEEexample:BSTcontrol}
\title{Low-Complexity ZF/MMSE  Receivers for MIMO-OTFS  Systems With Imperfect CSI}

\author{\IEEEauthorblockN{Prem Singh, Abhishek Gupta, Himanshu B. Mishra and Rohit Budhiraja} \thanks{Prem Singh, Abhishek Gupta and Rohit Budhiraja are with the Department of Electrical  Engineering, IIT Kanpur, 208016, India (e-mail: \{psrawat, gabhi, rohitbr\}$\emph{@}$iitk.ac.in).}
\thanks{Himanshu B. Mishra is with the Department of Electronics Engineering, IIT Dhanbad, India (e-mail: himanshu$\emph{@}$iitism.ac.in).  A part of this work is
submitted in IEEE International Conference on Communications (ICC) 2021.}
}

\IEEEtitleabstractindextext{

\begin{abstract}
Orthogonal time-frequency space (OTFS) scheme, which transforms a time and frequency selective channel into an almost non-selective channel in the delay-Doppler domain, establishes reliable wireless communication for high-speed moving devices. This work designs and analyzes low-complexity  zero-forcing (LZ) and minimum mean square error (LM) receivers for multiple-input multiple-output (MIMO)-OTFS systems with perfect and imperfect receive channel state information (CSI). The proposed receivers provide exactly the same solution as that of the conventional counterparts,  and reduce the complexity by exploiting the doubly-circulant nature of the MIMO-OTFS channel matrix, the block-wise inverse, and \emph{Schur} complement.  We also derive, by exploiting the Taylor expansion and results from random matrix theory, {a tight approximation of the post-processing signal-to-noise-plus-interference-ratio (SINR) expressions in closed-form for both LZ and LM receivers.} We show that the derived SINR expressions, when averaged over multiple channel realizations, accurately characterize their respective bit error rate (BER) of both perfect and imperfect receive CSI. We numerically show the lower BER and lower complexity of the proposed designs over state-of-the-art exiting solutions.
\end{abstract}

}

\maketitle

\IEEEdisplaynontitleabstractindextext

\IEEEpeerreviewmaketitle

\section{Introduction}
A practical wireless channel, due to multiple propagation paths and Doppler shift, is both time- and frequency-selective \cite{DBLP:books/cu/G2005}. Cyclic prefix (CP)-aided orthogonal frequency division multiplexing (OFDM)  is commonly used to combat the frequency-selectivity \cite{DBLP:books/cu/G2005}. The high Doppler shift, due to high-speed relative movement between transmitter and receiver, however, disturbs inter-subcarrier orthogonality in an OFDM system, which significantly degrades its performance  \cite{WangPMZ06}. The recently-proposed, orthogonal time frequency space (OTFS) scheme \cite{DBLP:journals/corr/abs-1802-02623, HadaniRTMGMC17} tackles this impairment by multiplexing transmit symbols in the delay-Doppler domain. This is  unlike the OFDM system which multiplexes them  in the  time-frequency domain.  

OTFS scheme uses inverse symplectic finite Fourier transform (ISFFT) to map transmit symbols in the delay-Doppler domain on to a set of two-dimensional time-frequency  orthogonal basis functions. This transform converts a  time- and frequency-selective channel for each transmit symbol in an OTFS frame into an almost flat-faded delay-Doppler channel.  This can be exploited for reducing both bit error rate (BER) and the pilot overhead required to estimate a rapidly time-varying channel. Furthermore, the delay-Doppler domain channel, due to small number of reflectors, is sparse  \cite{DBLP:books/ieee/Jakes74}, which can be leveraged to reduce channel estimation and data detection complexity \cite{RavitejaPH19,SurabhiC20}. For communication between vehicles with speeds ranging from $30$ km/h to $500$ km/h, OTFS scheme is shown to have significantly lower BER than the OFDM scheme~\cite{HadaniRTMGMC17}.

Designing computationally-efficient receivers for OTFS systems, which is also one of our objectives, has attracted significant research attention \cite{HadaniRTMGMC17,SurabhiAC19,raviteja2018interference, RavitejaPJHV18,RamachandranC18,SurabhiRC19,MuraliC18, SurabhiC20,TiwariDR19}. Hadani \textit{et al.} in \cite{HadaniRTMGMC17} numerically investigated the block error rate  of sphere decoding (SD) in single-input single-output (SISO)-OTFS systems. Reference \cite{SurabhiAC19} derived the diversity of SISO/multiple-input multiple-output (MIMO)-OTFS systems with  maximum-likelihood (ML) decoding. Raviteja \textit{et al.} in \cite{raviteja2018interference, RavitejaPJHV18}, by assuming perfect receive channel state information (CSI), proposed a reduced complexity iterative message passing (MP)-aided  data detection algorithm for SISO-OTFS systems. This algorithm exploits the inherent  OTFS channel sparsity by using sparse factor graphs. 

Ramachandran \textit{et al.} in \cite{RamachandranC18} investigated a MP receiver for MIMO-OTFS systems. Reference \cite{SurabhiRC19} numerically investigated the BER of an MP detector in mmWave  SISO-OTFS systems in  presence of phase noise.  Reference \cite{MuraliC18} modeled the equivalent channel matrix in vector form, and proposed a low-complexity Markov chain Monte Carlo sampling (MCMCS)  detector for SISO-OTFS systems.  Surabhi \textit{et al.} in \cite{SurabhiC20}, by exploiting the inherent OTFS circulant channel structure, proposed reduced complexity zero-forcing (ZF) and minimum mean square error (MMSE) receivers  for SISO-OTFS systems.  Tiwari \textit{et al.} in \cite{TiwariDR19}, by using sparsity and quasi-banded structure of matrices involved in the demodulation process, investigated a low-complexity linear MMSE (LMMSE) receiver for SISO-OTFS systems. Cheng \textit{et al.} in \cite{cheng2019low}, by using two-dimension fast Fourier transform, proposed reduced complexity ZF and MMSE equalizers for SISO-OTFS systems. The authors in \cite{surabhi2020low} developed low-complexity ZF/MMSE receivers for $2\times 2$ MIMO-OTFS systems. Table-\ref{table:review} summarizes the aforementioned low-complexity OTFS designs. 
 
  \begin{table}[htbp]
 	\centering
 	\caption{{Summary of literature focusing on data detection in OTFS}}
 	\begin{tabular}{| m{5.75em} | m{7em}| m{8em}| m{5em}| m{6.5em}|m{8.5em}|} 
 		\hline
 		Ref. & Scenario & CSI& Scheme& Analytical BER & Complexity Analysis\\ 
 		\hline
 		\cite{HadaniRTMGMC17} & SISO & Perfect & SD & $\times$ & $\times$ \\ 
 		\hline 
 		\cite{raviteja2018interference, RavitejaPJHV18} & SISO & Perfect & MP & $\times$ & \tick \\ 
 		\hline  
 		\cite{SurabhiRC19} & SISO & Perfect & MP & $\times$ & $\times$ \\ 
 		\hline 
 		\cite{MuraliC18} & SISO & Imperfect & MCMCS & $\times$ & $\times$ \\ 
 		\hline 
 		\cite{SurabhiC20,TiwariDR19,cheng2019low} & SISO & Perfect & ZF, MMSE & $\times$ & \tick \\ 
 		\hline 
 		\cite{surabhi2020low} & $2\times 2$ MIMO & Perfect & ZF, MMSE & $\times$ & \tick \\
 		\hline 
 		\cite{SurabhiAC19} & SISO, MIMO & Perfect & ML & $\times$ & $\times$ \\ 
 		\hline 
 		\cite{RamachandranC18} & MIMO & Perfect & MP & $\times$ & $\times$ \\ 
 		\hline 
 		\textbf{Proposed} & \textbf{MIMO} & \textbf{Perfect, Imperfect} & \textbf{ZF, MMSE} & \tick & \tick \\ 
 		\hline
 	\end{tabular}
 	\label{table:review}
 \end{table}

We see from Table-\ref{table:review} that the existing OTFS literature has not yet investigated computationally-efficient receivers for MIMO-OTFS systems, exception being the MP algorithm in \cite{RamachandranC18} and ZF/MMSE receivers for $2\times 2$ MIMO-OTFS systems in \cite{surabhi2020low}. Also these works, except \cite{MuraliC18}, consider perfect receive CSI, an assumption which is usually not applicable in practice.
Conventional ZF and MMSE receivers, due to inherent matrix inversion,  has $\mathcal{O}(N_t^3M^3N^3)$ complexity for MIMO-OTFS systems \cite{SurabhiC20}, where $N_t$, $M$ and $N$ denote the number of transmit antennas,  delay bins and Doppler bins, respectively. For practical systems, the parameters $M,N,N_t$ take large values, which radically increases the complexity of conventional ZF and MMSE receivers for MIMO-OTFS systems \cite{SurabhiC20,raviteja2018interference}. The MP receiver, which reduces complexity by using the Gaussian approximation of interference, and by exploiting sparsity in the OTFS channel, is  widely used  for MIMO-OTFS systems \cite{RavitejaPH19,RamachandranC18}. The MP receiver, however, as shown later in Section-\ref{Results}, has significantly higher complexity than the low-complexity designs proposed in this work. Moreover, the existing literature has not yet analytically derived BER for MIMO-OTFS systems with  imperfect receive CSI. In the light of above observations, this work focuses on designing low-complexity  receivers for MIMO-OTFS systems, and  deriving their analytical BER expressions. Towards achieving these aims, the \textbf{main contributions} of this work can be summarized as follows.
{\begin{itemize}
\item  The ZF and MMSE receivers invert the matrices $\mathbf{D}^H\mathbf{D}$ and $\mathbf{D}^H\mathbf{D}+\rho \mathbf{I}$, respectively. Here $\mathbf{D}\in \mathbb{C}^{N_rMN\times N_tMN}$, which  consists of eigenvalues of MIMO-OTFS channel matrix $\mathbf{H}$, unlike the SISO works in \cite{SurabhiC20,TiwariDR19,cheng2019low}, is a non-diagonal matrix. \textit{The challenge is to invert extremely large-dimensional non-diagonal matrices $\mathbf{D}^H\mathbf{D}$ and $\mathbf{D}^H\mathbf{D}+\rho \mathbf{I}$ in a computationally-efficient manner.} 
{We propose a novel computationally-efficient algorithm to invert  them by exploiting   i) the inherent doubly-circulant structure of $\mathbf{H}$,  block-wise inverse property of  block matrices and Schur complement; and ii) a recursive structure which iterates between matrix partitioning and backtracking phases.} To perform  block-wise inverse of a block matrix, we need to prove that its sub-matrix  and its \emph{Schur} complement are always invertible. We prove this by using intricate ideas from matrix theory.

\item  We use the proposed algorithm to construct low-complexity ZF (LZ) and MMSE (LM) receivers for MIMO-OTFS systems, {which yield exactly the same solution, and consequently the same BER}, as that of the conventional ZF and MMSE receivers for both perfect and imperfect receive CSI. The BER of  proposed receivers is further reduced by integrating them with likelihood ascent search (LAS) technique \cite{chockalingam2014large}.
We analytically and numerically show that the proposed LZ and LM receivers have significantly lower computational complexity than conventional ZF and MMSE receivers with $\mathcal{O}(N_t^3M^3N^3)$ complexity \cite{golub2012matrix}, and the MP-based scheme with $\mathcal{O}(N_I N_t N_r MNSQ)$ complexity  \cite{raviteja2018interference}. Here $N_I$ is the number of iterations required for convergence of MP algorithm, $Q$ is the number of constellation points, and $S$ is the number of non-zero elements in a row or column of OTFS channel matrix $\mathbf{H}_{r,t}\in\mathbb{C}^{MN\times MN}$ between the $t$th transmit and $r$th receive antenna.

\item We  derive tight closed-form approximate expressions for SINR of LM and LZ receivers, and use  the proposed algorithm for their low-complexity implementation. 
Recall that the proposed algorithm  computes low-complexity inverse of $\mathbf{D}_{\text{LM}}=\mathbf{D}^H\mathbf{D}+\rho \mathbf{I}$ and $\mathbf{D}_{\text{LZ}}=\mathbf{D}^H\mathbf{D}$ matrices. \textit{To use the algorithm, the challenge is to show that the SINR expressions indeed invert $\mathbf{D}_{\text{LM}}$/$\mathbf{D}_{\text{LZ}}$ matrix.} We show this by using 
i)  ideas from random matrix theory; ii) Taylor series expansion; and iii) properties of block matrices described in the sequel.

\item We numerically show that the  proposed designs, for both perfect and imperfect receive CSIs, have significantly lower BER and complexity than the existing designs  \cite{SurabhiAC19,raviteja2018interference,RamachandranC18}. We also show that the  BERs derived using the  SINR expressions, match the simulated ones.
\end{itemize}}
  
%\subsection{Organisation of the Paper}
%The next section describes the input output relation for  MIMO-OTFS system. Section-\ref{LCD} proposes the low-complexity ZF and MMSE receivers for MIMO-OTFS systems, followed by complexity evaluation of the proposed technique in  Section-\ref{complexity}. Section-\ref{BER} derives the SINR and BER for the proposed ZF and MMSE receivers in the presence of imperfect receive CSI. Section-\ref{Results} demonstrates our simulation results and Section-\ref{Conclusion} concludes the paper, followed by the supplementary proofs of various results in the appendices.
\textbf{Notations:}
Lower and upper case bold face letters $\mathbf{a}$ and $\mathbf{A}$ denote vectors and matrices. The superscript $(\cdot)^H$ and $(\cdot)^T$ denote Hermitian and transpose operators respectively,  and $\mathbf{A}\otimes \mathbf{B}$ denotes  Kronecker product of the matrices $\mathbf{A}$ and~$\mathbf{B}$. The operator $\text{vec}(\mathbf{A})$ vectorises the matrix $\mathbf{A}$ and the operator $\mathbb{E}[\cdot]$ represents the expectation of a random variable. The operation $[\mathbf{A}]_{(p,q)}$ extracts the $(p,q)$th element of the matrix $\mathbf{A}$, and $[\mathbf{D}]_{p,q}$ extracts the $(p,q)$th block of the matrix $\mathbf{D}$. The notations $\text{diag}[a_1,a_2,\ldots,a_{N}]$ and $\mbox{blkdiag}[\mathbf{A}_1,\mathbf{A}_2,\ldots,\mathbf{A}_N]$ denote a diagonal matrix and  a block diagonal matrix, respectively. The notations $\mathbf{I}_{N}$,  $\mathbf{0}_{M\times N}$ and $[m-n]_{M}$ represent an $N\times N$ identity matrix, $M\times N$ zero matrix and modulo-$M$ operation, respectively. %The symbol $\mathcal{{B}}_{M,N}$ denotes a set of circulant matrices with $M$ circulant blocks, each of size $N\times N$. The notation $\mathcal{C}_{N_rN_t,MN}$ denotes the set of block matrices with $N_rN_t$ blocks of $MN\times MN$ diagonal matrices, and $\mathcal{C}_{t^2,MN}$ represents the sets of block square matrices with $t^2$ blocks of $MN\times MN$ diagonal matrices. 
%The notation $X\sim\mathcal{CN}(0,\sigma^2)$ denotes a zero-mean circularly symmetric complex Gaussian random variable $X$ with variance $\sigma^2$. 



\section{MIMO-OTFS system model}
\label{system_MIMO}
For a better understanding of MIMO-OTFS system model, we first explain the SISO-OTFS system model, and then extend it to its MIMO counterpart. We begin by considering an OTFS frame in the delay-Doppler domain with $ N $ Doppler bins and $ M $ delay bins \cite{raviteja2018interference}. Let $ x\left[k,l\right] $ be a QAM symbol, which is to be transmitted on the $ k $th Doppler and the $ l $th delay bin in the delay-Doppler frame, where $ k=0,1,\ldots,N-1 $ and $ l=0,1, \ldots,M-1$. The $MN$ symbols in the delay-Doppler domain in the OTFS scheme are first transformed to $MN$ symbols in the time-frequency domain by using  ISFFT. The transformed symbol at the time index $n$ and the frequency index $m$, for $0\leq m\leq M-1$ and $0\leq n\leq N-1$, is given as follows \cite{raviteja2018interference}
\begin{equation}
\nonumber x^{\text{TF}}\left[n,m\right]=\dfrac{1}{MN}\sum_{k=0}^{N-1}\sum_{l=0}^{M-1}x\left[k,l\right]\exp\left\{j2\pi\Big(\dfrac{nk}{N}-\dfrac{ml}{M} \Big)\right\}.
\end{equation}
With the subcarrier spacing of $ \Delta f=1/T $, the time-frequency frame has a duration of $ NT $ and a bandwidth of $ M\Delta f $ \cite{raviteja2018interference}. The time-frequency domain symbols are then pulse-shaped using the Heisenberg transformation to yield the time domain signal $x(t)$ as \cite{raviteja2018interference}
\begin{equation}
\nonumber x(t)=\sum_{n=0}^{N-1}\sum_{m=0}^{M-1}x^{\text{TF}}[n,m]p_{\text{tx}}(t-nT)e^{j2\pi m\Delta f(t-nT)}.
\end{equation}
Here $p_{\text{tx}}(t)$ is the impulse response of the pulse-shaping filter. The time-domain transmit signal $x(t)$, after passing through a time-varying wireless channel, is received as \cite{raviteja2018interference}
\begin{equation}
\nonumber y(t)=\int_{\nu}\int_{\tau}h\left(\tau,\nu\right)x(t-\tau)d\tau d\nu.
\end{equation}
The delay-Doppler domain wireless channel $h\left(\tau,\nu\right)$, with delay parameter $\tau$ and Doppler parameter $\nu$, is \cite{raviteja2018interference} 
\begin{equation}\label{eq:DD_channel}
h\left(\tau,\nu\right)=\sum_{i=1}^{L_{h}}h_i\delta\left(\tau-\tau_{i} \right)\delta\left(\nu-\nu_{i} \right).
\end{equation}  
Here $ L_{h} $ is the number of channel paths due to $L_h$ clusters of reflectors, where each cluster introduces a delay and a Doppler shift \cite{RavitejaPJHV18}. The three-tuple  $ (h_{i},\tau_{i},  \nu_{i})$ denotes complex channel gain $h_i\sim\mathcal{CN}(0,\sigma^2_{L_{h_{i}}})$ \cite{SurabhiAC19}, delay $ \tau_{i} $ and the Doppler $ \nu_{i} $ for the $i$th cluster.  The delay and Doppler taps for the $i$th path are obtained as   
\begin{equation}
\tau_{i}=\frac{l_{i}}{M\Delta f}, \;\nu_{i}=\frac{k_{i}}{NT},
\end{equation}
where $ l_{i} $ and $ k_{i} $ are  integer indices corresponding to the delay $ \tau_{i} $ and Doppler $ \nu_{i} $, respectively. Both delay and Doppler values need not be integer multiple of taps $l_i$ and $k_i$ \cite{SurabhiAC19}; their discretization, however,  allows us to model the channel with fewer delay and Doppler taps \cite{FishGHSS13}.

The receiver, on its received signal $y(t)$,  performs Wigner transform \cite{SurabhiAC19}, which uses a receive pulse shape which  is matched to the transmit pulse shape $p_{\text{tx}}(t)$. {We consider, as commonly assumed  in the literature \cite{SurabhiC20,SurabhiAC19}, a MIMO OTFS system with ideal waveform, which satisfies bi-orthogonality and robustness conditions \cite{HadaniRTMGMC17}. The proposed receivers thus lower bound the performance of OTFS system with practically realizable waveforms.}  
The time-frequency domain signal, after sampling Wigner transformed signal at $t=nT$ and $f=m\Delta f$, is given as \cite{SurabhiAC19}
\begin{align}
y^{\text{TF}}[n,m]=H[n,m]x^{\text{TF}}[n,m]+V[n,m]. 
\end{align}
Here $V[n,m]$ is the additive white Gaussian noise at the output of Wigner transform, and %$H[n,m]$ is \cite{SurabhiAC19}
\begin{equation}
\nonumber H[n,m]=\int_{\nu}\int_{\tau}h\left(\tau,\nu\right)e^{j2\pi\nu nT}e^{-j2\pi(\nu+m\Delta f)\tau}d\tau d\nu.
\end{equation}
Finally, the time-frequency signal $y^{\text{TF}}[n,m]$ is transformed to the delay-Doppler domain using SFFT as follows \cite{raviteja2018interference}
\begin{equation}
\nonumber y\left[k,l\right]=\sum_{n=0}^{N-1}\sum_{m=0}^{M-1}y^{\text{TF}}[n,m]\exp\left\{-j2\pi\Big(\dfrac{nk}{N}-\dfrac{ml}{M} \Big)\right\}.
\end{equation}
By substituting the expression for $y^{\text{TF}}[n,m]$ and $x^{\text{TF}}[n,m]$ in the above expression, the delay-Doppler domain receive signal $y[k,l]$ is re-expressed as follows \cite{SurabhiAC19}
\begin{equation}
\nonumber y\left[k,l\right]=\sum_{\bar{k}=0}^{N-1}\sum_{\bar{l}=0}^{M-1}x\big[\bar{k},\bar{l}\big]h_{c}\left(\dfrac{k-\bar{k}}{NT},\dfrac{l-\bar{l}}{M\Delta f}\right)+v[k,l].
\end{equation}
Here $h_c(\bar{\nu},\bar{\tau})$, with $\bar{\nu}=(k-\bar{k})/(NT)$ and $\bar{\tau}=(l-\bar{l})/(M\Delta f)$, is the output of circular convolution between channel $h(\nu,\tau)$ and the windowing function $w(\nu,\tau)$, and is given as \cite{SurabhiAC19} 
\begin{equation}
\nonumber h_c(\bar{\nu},\bar{\tau})= \int_{\nu}\int_{\tau}h\left(\tau,\nu\right)w\left(\bar{\nu}-\nu,\bar{\tau}-\tau\right)d\tau d\nu.
\end{equation}
The windowing function $w(\bar{\nu},\bar{\tau})$ is calculated using the SSFT of transmit and receive time-frequency windowing functions $W_{\text{tx}}[n,m]$ and $W_{\text{rx}}[n,m]$ as \cite{HadaniRTMGMC17}
\begin{equation}
\nonumber w(\bar{\nu},\bar{\tau})= \sum_{n=0}^{N-1}\sum_{m=0}^{M-1}W_{\text{tx}}[n,m]W_{\text{rx}}[n,m]e^{-j2\pi(\bar{\nu} nT-\bar{\tau} m\Delta f)}.
\end{equation}
By  assuming ideal transmit and receive windowing functions \cite{SurabhiAC19}, the input-output relation for the channel model in \eqref{eq:DD_channel}, similar to \cite{RavitejaPJHV18}, can be expressed as follows
\begin{equation}\label{eq:in_out}
y\left[k,l \right]=\sum_{i=1}^{L_{h}}{h}'_{i}x\left[\left(k-k_{i} \right)_{N},\left(l-l_{i} \right)_{M} \right]+v\left[k,l \right].
\end{equation}
Here $ v\left[k,l \right] $ is the circularly symmetric zero mean complex Gaussian noise with variance $ \sigma_{v}^{2} $, and $ {h}'_{i}=h_{i}\exp\left(-j2\pi \nu_{i}\tau_{i} \right) $. Note that since $h_i\sim\mathcal{CN}(0,\sigma^2_{h_{i}})$, so is ${h}'_{i}$. 
The SISO-OTFS receive signal in \eqref{eq:in_out}, for mathematical brevity, is expressed in vector form as follows \cite{SurabhiC20,DingSFP19a} %\colr{see notations}
\begin{equation}
\label{eqn:rx_siso}
\mathbf{y}_{\text{s}}=\mathbf{H}_{\text{s}}\mathbf{x}_{\text{s}}+\mathbf{v}_{\text{s}},
\end{equation}
where, for $ k=0,\ldots,N-1$ and $ l=0,\ldots,M-1 $, the $ \left(k+Nl\right) $th element of $ \mathbf{x}_{\text{s}}\in \mathbb{C}^{MN\times 1} $ is $ x_{k+Nl}=x\left[k,l \right]$. The received vector $ \mathbf{y}_{\text{s}}\in \mathbb{C}^{MN \times 1} $ and noise vector $ \mathbf{v}_{\text{s}}\in \mathbb{C}^{MN \times 1} $ also have the same structure. %\colr{Sir, there is no fix structure for defining (5) mathematically since their structure depends on the delay-Doppler locations. Some authors have taken examples to define OTFS model. If you want, we can also add one example for the same. Others are citing references for their system model. I have also added references [7,19] to clarify this}.
\begin{observation}
 Here $\mathbf{H}_{\text{s}}\in \mathbb{C}^{MN \times MN} $ is a doubly-block circulant matrix, i.e., it is a block circulant matrix with $M$ circulant blocks, each of size $N\times N$ \cite{DingSFP19a}. 
\end{observation}
The SISO-OTFS system model in \eqref{eqn:rx_siso} can now be extended to the MIMO scenario.  We consider a spatially-multiplexed MIMO-OTFS system with $N_t$ transmit antennas transmits $N_t$ data streams, and has  $N_r\geq N_t$ receive antennas.   The signal at the $r$th receive antenna, using \eqref{eqn:rx_siso}, is  given as 
\begin{equation}\label{eq:mimo1}
\mathbf{y}_r = \sum_{t=1}^{N_t}\mathbf{H}_{r,t}\mathbf{x}_t+\tilde{\mathbf{v}}_r.
\end{equation}
Here $\mathbf{x}_t\in\mathbb{C}^{NM\times 1}$ and $\mathbf{y}_r\in\mathbb{C}^{NM\times 1}$ are transmit  and receive vectors for the $t$th transmit and $r$th receive antenna, respectively. Each element of $\mathbf{x}_t$ is independent and identically distributed (i.i.d.) with mean zero and variance $P_x$ \cite{DBLP:books/cu/G2005}. Each element of the noise vector $\tilde{\mathbf{v}}_r\in\mathbb{C}^{NM\times 1}$ at the $r$th receive antenna follows circularly-symmetric complex Gaussian distribution with mean zero and variance $\sigma^2_v$, given as $\mathcal{CN}(0,\sigma^2_v)$. The matrix $\mathbf{H}_{r,t}\in\mathbb{C}^{NM\times NM}$ represents the OTFS channel in the delay-Doppler domain between the $t$th transmit and the $r$th receive antenna. 
\begin{observation}
\label{obs_mimo_ch1}
The matrix $\mathbf{H}_{r,t}$, similar to $\mathbf{H}_{\text{s}}$,  is a doubly-block circulant matrix, i.e., it is a block circulant matrix with $M$ circulant blocks, each of size $N\times N$ \cite{raviteja2018interference}. 
\end{observation}

We concatenate receive vectors in \eqref{eq:mimo1} as $\mathbf{y}=[\mathbf{y}_1^T,\mathbf{y}_2^T,\ldots,\mathbf{y}_{N_r}^T]^T\in\mathbb{C}^{N_r MN\times 1}$ to obtain
\begin{equation}\label{eq:mimo_otfs}
\mathbf{y} = \mathbf{H}\mathbf{x}+\tilde{\mathbf{v}}.
\end{equation}
Here $\mathbf{x}=[\mathbf{x}_1^T,\mathbf{x}_2^T,\ldots,\mathbf{x}_{N_t}^T]^T\in\mathbb{C}^{N_t MN\times 1}$ is transmit symbol vector and  $\tilde{\mathbf{v}}=[\tilde{\mathbf{v}}_1^T,\tilde{\mathbf{v}}_2^T,\ldots,\tilde{\mathbf{v}}_{N_r}^T]^T\in\mathbb{C}^{N_r MN\times 1}$, with probability density function (pdf) $\mathcal{CN}(\mathbf{0},\sigma^2_v\mathbf{I}_{N_r MN})$,  denotes noise vector. The block channel matrix $\mathbf{H}\in\mathbb{C}^{N_r MN\times N_t MN}$ for MIMO-OTFS system in \eqref{eq:mimo_otfs} is given as follows 
\begin{equation}\label{eq:mimo_otfs_mtx}
\mathbf{H} \overset{\Delta}{=} \begin{bmatrix}
\mathbf{H}_{1,1} & \mathbf{H}_{1,2} & \cdots & \mathbf{H}_{1,N_t} \\ 
\mathbf{H}_{2,1} & \mathbf{H}_{2,2} & \cdots & \mathbf{H}_{2,N_t} \\ 
\vdots & \vdots & \ddots & \vdots \\ 
\mathbf{H}_{N_r,1} & \mathbf{H}_{N_r,2} & \cdots & \mathbf{H}_{N_r,N_t}  
\end{bmatrix}.
\end{equation}
Let $\mathbf{G}_{\text{A}}\in\mathbb{C}^{N_rMN\times N_tMN}$, with $\text{A}\in\{\text{ZF}, \text{MMSE}\}$, be the receiver matrix for detecting the transmit vector $\mathbf{x}$, whose estimate is $\hat{\mathbf{x}}=\mathbf{G}^H_{\text{A}}\mathbf{y}$. For the conventional ZF and MMSE receivers, the matrix $\mathbf{G}_{\text{A}}$ can be expressed as \cite{SinghMJVH20}
\begin{eqnarray}\label{eq:Combiners}
\mathbf{G}_{\text{A}}=
 \left\{
  \begin{array}{@{}ll@{}}
   \mathbf{H}\big(\mathbf{H}^{H}\mathbf{H}\big)^{-1}& \text{for}\ \text{ZF}\\
   \mathbf{H}\Big(\mathbf{H}^H\mathbf{H}+\rho\mathbf{I}_{N_tMN}\Big)^{-1}& \text{for}\ \text{MMSE},
  \end{array}\right.
\end{eqnarray}
where $\rho={\sigma^{2}_{v}}/{P_x}$ = 1/SNR. These receivers invert a matrix of size $N_tMN\times N_tMN$ with $\mathcal{O}(N^3_tM^3N^3)$ complexity. The number of delay bins $M$, Doppler bins $N$ and transmit antennas $N_t$ for practical systems take large values \cite{RavitejaPJHV18}. The  conventional ZF and MMSE receivers for MIMO-OTFS systems are thus computationally inefficient. We  propose low-complexity ZF (LZ)  and MMSE (LM) receivers, which exploit the inherent structure of the MIMO-OTFS channel matrix $\mathbf{H}$, and  have significantly-reduced $\mathcal{O}(MN)+\mathcal{O}(MN\mbox{log}_2MN)$ complexity. 

\section{Proposed low-complexity receivers for MIMO-OTFS systems}
\label{LCD}
The proposed LZ and LM receivers use the following key ideas: 
a) split MIMO-OTFS channel $\mathbf{H}$ in terms of  DFT and diagonal block matrices by exploiting its inherent circulant structure; and b) invoke block-wise inverse property for inverting the matrix in the MMSE and ZF receivers.

\textbf{{Preliminaries:}}
Before discussing the proposed receivers, we define in Table~\ref{table:sets}, various sets of matrices which will be used frequently in the sequel. 
\begin{table}[h!]
	\centering
	\caption{{Definition of different sets used in the work.}}
	\begin{tabular}{| m{5em} | m{32em}|} 
		\hline
		\textbf{Set} & \textbf{Definition}  \\ 
		\hline
		$ \mathcal{B}_{M,N}$ & Set of circulant matrices with $M$ circulant blocks, each of size $N\times N$. \\ 
		\hline
		$ \mathcal{C}_{N_rN_t,MN}$ & Set of block matrices with $N_rN_t$ blocks of $MN\times MN$ diagonal matrices  \\
		\hline 
		$ \mathcal{C}_{t^2,MN}$ & Set of block square matrices with $t^2$ blocks of $MN\times MN$ diagonal  matrices \\
		\hline 
	\end{tabular}
	\label{table:sets}
\end{table}
We see from Table-\ref{table:sets} that the set $\mathcal{{B}}_{M,N}$ denotes a set of circulant matrices with $M$ circulant blocks, each of size $N\times N$. If a matrix $\mathbf{B}\in\mathcal{B}_{M,N}$, it can be represented as $\mathbf{B}=\mathtt{CIRC}(\mathbf{B}_1,\mathbf{B}_2,\ldots,\mathbf{B}_{M})$, where  $\mathtt{CIRC}(\cdot)$ is circulant operation and $\mathbf{B}_i\in\mathbb{C}^{N\times N}$ is the $i$th circulant block of $\mathbf{B}$. We now state a lemma from \cite{kra2012circulant}. 
{\begin{lemma}\label{Lemma0}
If a matrix $\mathbf{B}\in\mathcal{B}_{M,N}$, it can be diagonalized using the DFT matrices $\mathbf{F}_M\in\mathbb{C}^{M\times M}$ and $\mathbf{F}_N\in\mathbb{C}^{N\times N}$ as $\mathbf{B}=(\mathbf{F}_M\otimes\mathbf{F}_N)^H\boldsymbol{\Lambda}(\mathbf{F}_M\otimes\mathbf{F}_N)$. The diagonal matrix $\boldsymbol{\Lambda}\in\mathbb{C}^{MN\times MN}$, which consists of the eigenvalues $\lambda_1,\lambda_2,\ldots,\lambda_{MN}$ of the matrix $\mathbf{B}$, can be expressed as
		\begin{equation}\label{eq:Prop2}
		\boldsymbol{\Lambda}=\sum_{i=0}^{M-1}\boldsymbol{\Omega}_{M}^i\otimes \boldsymbol{\Lambda}^i.
		\end{equation}
The diagonal matrix $\boldsymbol{\Lambda}^i\in\mathbb{C}^{N\times N}$ consists of the eigenvalues of the $i$th circulant block $\mathbf{B}_i$ and $\boldsymbol{\Omega}_{M}=\mbox{diag}[1,e^{j2\pi/M},e^{j4\pi/M},\ldots,e^{j2\pi(M-1)/M}]$.
\end{lemma}}
% It follows from \cite{8918014} that for any two matrices $\mathbf{X},\mathbf{Y}\in\mathcal{B}_{M,N}$, all the matrices obtained using the operations $\mathbf{X}^T$, $\mathbf{X}^{H}$, $\mathbf{X}^{-1}$ (if exist), $\mathbf{XY}$ and $a\mathbf{X}+b\mathbf{Y}$ (for arbitrary constants $a$ and $b$) also belong to the set $\mathcal{B}_{M,N}$.

We recall from \textit{Observation~\ref{obs_mimo_ch1}} that each sub-matrix of the MIMO-OTFS channel matrix $\mathbf{H}$ in \eqref{eq:mimo_otfs_mtx} belongs to the set $\mathcal{B}_{M,N}$.  Using \emph{Lemma} \ref{Lemma0}, the $(r,t)$th sub-matrix $\mathbf{H}_{r,t}\in\mathbb{C}^{MN\times MN}$ of $\mathbf{H}$ can, therefore, be decomposed as $\mathbf{H}_{r,t}=(\mathbf{F}_M\otimes\mathbf{F}_N)^H\mathbf{D}_{r,t}(\mathbf{F}_M\otimes\mathbf{F}_N)$, where the diagonal matrix $\mathbf{D}_{r,t}=\mbox{diag}[\lambda^{r,t}_1,\lambda^{r,t}_2,\ldots,\lambda^{r,t}_{MN}]\in\mathbb{C}^{MN\times MN}$ consists of eigenvalues of $\mathbf{H}_{r,t}$. From \eqref{eq:Prop2}, we have $\mathbf{D}_{r,t}=\sum_{i=1}^{M}\boldsymbol{\Omega}_{M}^i\otimes\mathbf{D}^i_{r,t}$, where diagonal matrix $\mathbf{D}^i_{r,t}$ consists  of eigenvalues of the $i$th circulant block of $\mathbf{H}_{r,t}$. The MIMO-OTFS channel in \eqref{eq:mimo_otfs_mtx}, using this decomposition,  can be partitioned as
\begin{equation}\label{eq:mimo_otfs_mtx_D}
\mathbf{H} = \boldsymbol{\Psi}_{\text{R}}^H\mathbf{D}\boldsymbol{\Psi}_{\text{T}}.
\end{equation}
Here $\boldsymbol{\Psi}_{\text{R}}\in\mathbb{C}^{N_rMN\times N_rMN}=\mathbf{I}_{N_r}\otimes\mathbf{F}_M\otimes\mathbf{F}_N$ and $\boldsymbol{\Psi}_{\text{T}}\in\mathbb{C}^{N_tMN\times N_tMN}=\mathbf{I}_{N_t}\otimes\mathbf{F}_M\otimes\mathbf{F}_N$.
 Since $\mathbf{F}_{M}$ and $\mathbf{F}_{N}$ are DFT matrices, $\boldsymbol{\Psi}_{\text{R}}\boldsymbol{\Psi}_{\text{R}}^{H}=\boldsymbol{\Psi}^{H}_{\text{R}}\boldsymbol{\Psi}_{\text{R}}=\mathbf{I}_{N_rMN}$ and $\boldsymbol{\Psi}_{\text{T}}\boldsymbol{\Psi}_{\text{T}}^{H}=\boldsymbol{\Psi}^{H}_{\text{T}}\boldsymbol{\Psi}_{\text{T}}=\mathbf{I}_{N_tMN}$. The block matrix $\mathbf{D}\in\mathbb{C}^{N_rMN\times N_tMN}$, which consists of eigenvalues of $\mathbf{H}$, is given as
\begin{equation}\label{eq:mimo_otfs_mtx_D1}
\mathbf{D} = \begin{bmatrix}
\mathbf{D}_{1,1} & \mathbf{D}_{1,2} & \cdots & \mathbf{D}_{1,N_t} \\ 
\mathbf{D}_{2,1} & \mathbf{D}_{2,2} & \cdots & \mathbf{D}_{2,N_t} \\ 
\vdots & \vdots & \ddots & \vdots \\ 
\mathbf{D}_{N_r,1} & \mathbf{D}_{N_r,2} & \cdots & \mathbf{D}_{N_r,N_t}  
\end{bmatrix}.
\end{equation}
\textit{We see that the eigenvalue matrix $\mathbf{D}$, unlike its counterpart in SISO-OTFS channel matrix,  is not diagonal, which is a key difference between the SISO- and MIMO-OTFS system models.} We also note that due to highly time-varying nature of channel, MIMO-OTFS and MIMO-OFDM system models are completely different \cite{RottenbergMHL19}.  Recalling the definition of the set $\mathcal{C}_{N_rN_t,MN}$ from Table \ref{table:sets}, we see that the matrix $\mathbf{D}$ belongs to the set $\mathcal{C}_{N_rN_t,MN}$. Using the decomposition in  
\eqref{eq:mimo_otfs_mtx_D}, and the properties $\boldsymbol{\Psi}_{\text{R}}\boldsymbol{\Psi}_{\text{R}}^{H}=\boldsymbol{\Psi}^{H}_{\text{R}}\boldsymbol{\Psi}_{\text{R}}=\mathbf{I}_{N_rMN}$ and $\boldsymbol{\Psi}_{\text{T}}\boldsymbol{\Psi}_{\text{T}}^{H}=\boldsymbol{\Psi}^{H}_{\text{T}}\boldsymbol{\Psi}_{\text{T}}=\mathbf{I}_{N_tMN}$, the equivalent combiner matrices for the proposed LZ and LM receivers are derived from \eqref{eq:Combiners} as follows
\begin{eqnarray}\label{eq:Combiners1}
\mathbf{G}_{\text{A}}=
 \left\{
  \begin{array}{@{}ll@{}}
   \boldsymbol{\Psi}^H_\text{R}\mathbf{D}\big(\mathbf{D}^{H}\mathbf{D}\big)^{-1}\boldsymbol{\Psi}_\text{T}& \text{for}\ \text{LZ}\\
   \boldsymbol{\Psi}^H_\text{R}\mathbf{D}\left(\mathbf{D}^H\mathbf{D}+\rho\mathbf{I}_{N_tMN}\right)^{-1}\boldsymbol{\Psi}_\text{T}& \text{for}\ \text{LM}.
  \end{array}\right.
\end{eqnarray}

\begin{observation}\label{Obser_cnv_lc}
The proposed LZ and LM receivers in \eqref{eq:Combiners1} yield exactly the same solutions as that of the conventional ZF and MMSE receivers in \eqref{eq:Combiners}. This is because, to derive \eqref{eq:Combiners1} from \eqref{eq:Combiners}, we replace the MIMO-OTFS channel matrix $\mathbf{H}$ by its decomposition in \eqref{eq:mimo_otfs_mtx_D}.
\end{observation}
For the sets $\mathcal{C}_{N_rN_t,MN}$ and $\mathcal{C}_{t^2,MN}$ in Table-\ref{table:sets}, we now state the following lemma from \cite{petersen2012matrix}.
\begin{lemma}\label{Lemma1}
 If $\mathbf{X}$, $\mathbf{Y}\in\mathcal{C}_{N_rN_t,MN}$, the matrices obtained using operations $\mathbf{X}^{T}$, $\mathbf{X}^H$, $\mathbf{XY}(=\mathbf{YX})$, $a_1\mathbf{X}+a_2\mathbf{Y}$ and $\sum_{n=1}^{N}a_n\mathbf{X}_n$ also belong to the set $\mathcal{C}_{N_rN_t,MN}$, where $a_1,a_2,\ldots,a_n$ are scalars. Additionally, if $\mathbf{X}$, $\mathbf{Y}\in\mathbb{C}^{N_rMN\times N_tMN}$, the matrices $\mathbf{X}\mathbf{Y}^H$ and $\mathbf{X}^H\mathbf{Y}$ belong to the set $\mathcal{C}_{N_r^2,MN}$ and $\mathcal{C}_{N_t^2,MN}$, respectively.
\end{lemma}

We see from \eqref{eq:Combiners} that the conventional ZF and MMSE receivers invert matrices $\mathbf{H}^H\mathbf{H}$ and $\mathbf{H}^H\mathbf{H}+\rho\mathbf{I}_{N_tMN}$ respectively with $\mathcal{O}(N_t^3M^3N^3)$ complexity \cite{golub2012matrix}. The proposed LZ and LM receivers use the equivalent formulation in \eqref{eq:Combiners1}, and therefore, invert  $\mathbf{D}_{\text{LZ}}=\mathbf{D}^H\mathbf{D}$ and $\mathbf{D}_{\text{LM}}=\mathbf{D}^H\mathbf{D}+\rho\mathbf{I}_{N_tMN}$ respectively.  By exploiting the properties of sets $\mathcal{C}_{N_rN_t,MN}$ and $\mathcal{C}_{t^2,MN}$, and \emph{Lemma} \ref{Lemma1}, we now propose a low-complexity algorithm for computing $\mathbf{D}_{\text{LZ}}^{-1}$ and $\mathbf{D}_{\text{LM}}^{-1}$, which as shown later in this section, has $\mathcal{O}(MN)$ complexity.
%\begin{enumerate}[label=P{{\arabic*}})]
%    \item
%    If $\mathbf{X}$, $\mathbf{Y}\in\mathcal{C}_{N_rN_t,MN}$, it can be readily verified that the matrices obtained by the operations $\mathbf{X}^{T}$, $\mathbf{Y}^H$, $\mathbf{XY}(=\mathbf{YX})$, $a\mathbf{X}+b\mathbf{Y}$ with constants $a$ and $b$, and $\sum_{n=1}^{N}a_n\mathbf{X}_n$ also belong to the set $\mathcal{C}_{N_rN_t,MN}$.
%    \item
%   If $\mathbf{X}\in\mathcal{C}_{N_rN_t,MN}$, then the matrix $\mathbf{X}^H\mathbf{X}\in\mathcal{C}_{N^2_t,MN}$, which is a set of square matrices of size $N_tMN\times N_t MN$ with $N^2_t$ blocks of diagonal matrices, each of size $MN\times MN$. The inverse $(\mathbf{X}^H\mathbf{X})^{-1}$ also belongs to the set $\mathcal{C}_{N^2_t,MN}$. The proof of this result is given in the Appendix-??.
%\end{enumerate} Let $\mathcal{C}_{t^2,MN}$, for $1\leq t\leq N_t$, be the set of block square matrices with $t^2$ blocks of diagonal  matrices of size $MN\times MN$.
\subsection{{Proposed low-complexity ZF (LZ) and MMSE (LM) receivers}}
 It follows from \emph{Lemma} \ref{Lemma1} that the matrix $\mathbf{D}_{\text{A}}\in\mathbb{C}^{N_tMN\times N_t MN}$, for $\text{A}\in\{\text{LZ, LM}\}$, belongs to the set $\mathcal{C}_{t^2,MN}$, where $t=N_t$. The matrix $\mathbf{D}_{\text{A}}$, thus consists of blocks of $MN\times MN$ diagonal matrices, and therefore, it can always be partitioned as follows
\begin{align}\label{eq:D_A}
\setlength{\arraycolsep}{2.5pt} % default: 5pt
\medmuskip = 1mu
\mathbf{D}_{\text{A}} =\begin{bmatrix}
\begin{array}{@{}cccc|c@{}}
\mathbf{D}_{\text{A}_{1,1}} & \mathbf{D}_{\text{A}_{1,2}} & \cdots & \mathbf{D}_{\text{A}_{1,N_t-1}}& \mathbf{D}_{\text{A}_{1,N_t}} \\ 
\mathbf{D}_{\text{A}_{2,1}} & \mathbf{D}_{\text{A}_{2,2}} & \cdots & \mathbf{D}_{\text{A}_{2,N_t-1}} & \mathbf{D}_{\text{A}_{2,N_t}} \\ 
\vdots & \vdots & \ddots &  \vdots &\vdots \\
\mathbf{D}_{\text{A}_{N_t-1,1}} & \mathbf{D}_{\text{A}_{N_t-1,2}} & \cdots & \mathbf{D}_{\text{A}_{N_t-1,N_t-1}} & \mathbf{D}_{\text{A}_{N_t-1,N_t}} \\  \hline
\mathbf{D}_{\text{A}_{N_t,1}} & \mathbf{D}_{\text{A}_{N_t,2}} & \cdots & \mathbf{D}_{\text{A}_{N_t,N_t-1}} & \mathbf{D}_{\text{A}_{N_t,N_t}}
  \end{array}
  \end{bmatrix}.
\end{align}
Here each $\mathbf{D}_{\text{A}_{i,j}}$, for $1\leq i,j\leq N_t$, is a diagonal matrix of size $MN\times MN$. If a matrix $\mathbf{X}$ can be partitioned in to four sub-matrices $\mathbf{A}$, $\mathbf{B}$, $\mathbf{C}$ and $\mathbf{D}$, it can be inverted block-wise as \cite{lu2002inverses}: 
\begin{equation}\label{eq:mimo_otfs_mtx1}
\setlength{\arraycolsep}{2.5pt} % default: 5pt
\medmuskip = 1mu
 \begin{bmatrix}
\mathbf{A}& \mathbf{B}\\ 
\mathbf{C}& \mathbf{D}  
\end{bmatrix}^{-1}=\begin{bmatrix}\mathbf{S}^{-1}& -\mathbf{S}^{-1}\mathbf{B}\mathbf{D}^{-1}\\ 
-\mathbf{D}^{-1}\mathbf{C}\mathbf{S}^{-1}& \mathbf{D}^{-1}+\mathbf{D}^{-1}\mathbf{C}\mathbf{S}^{-1}\mathbf{B}\mathbf{D}^{-1} 
\end{bmatrix}.
\end{equation}
This holds provided the matrix $\mathbf{D}$ and its \emph{Schur} complement $\mathbf{S}=\mathbf{A}-\mathbf{B}\mathbf{D}^{-1}\mathbf{C}$ are invertible. 
\begin{observation}
We observe from \eqref{eq:D_A} and \eqref{eq:mimo_otfs_mtx1} that the inverse of the matrix $\mathbf{D}_{\text{A}}$ can be performed block-wise, provided sub-matrix $\mathbf{D}_{\text{A}_{N_t,N_t}}$ in \eqref{eq:D_A} and its \emph{Schur} complement are always invertible. \emph{This is a key design aspect of the proposed low-complexity receivers for MIMO-OTFS systems.} {To perform block-wise inverse of the matrix $\mathbf{D}_{\text{A}}$ using \eqref{eq:mimo_otfs_mtx1} for reducing complexity, we next prove that the sub-matrix $\mathbf{D}_{\text{A}_{N_t,N_t}}$ in \eqref{eq:D_A} and its \emph{Schur} complement are always invertible. To this end, we state the next lemma whose proof is relegated to Appendix~\ref{Lemma2_proof}.}
\end{observation}
%\begin{proposition}\label{P1}
%If a matrix $\mathbf{X}$ can be partitioned in to four sub-matrices $\mathbf{A}$, $\mathbf{B}$, $\mathbf{C}$ and $\mathbf{D}$, its inverse can be performed block-wise as
%\begin{equation}\label{eq:mimo_otfs_mtx}
% \begin{bmatrix}
%\mathbf{A}& \mathbf{B}\\ 
%\mathbf{C}& \mathbf{D}  
%\end{bmatrix}^{-1}=\begin{bmatrix}\mathbf{S}^{-1}& -\mathbf{S}^{-1}\mathbf{B}\mathbf{D}^{-1}\\ 
%-\mathbf{D}^{-1}\mathbf{C}\mathbf{S}^{-1}& \mathbf{D}^{-1}+\mathbf{D}^{-1}\mathbf{C}\mathbf{S}^{-1}\mathbf{B}\mathbf{D}^{-1} 
%\end{bmatrix},
%\end{equation}
%provided the matrix $\mathbf{D}$ and its \emph{Schur} complement $\mathbf{S}=\mathbf{A}-\mathbf{B}\mathbf{D}^{-1}\mathbf{C}$ are invertible.
%\end{proposition}
%\begin{IEEEproof}
%Refer to \cite{lu2002inverses}.
%\end{IEEEproof}
\begin{lemma}\label{Lemma2}
If a matrix $\mathbf{X}\in\mathcal{C}_{t^2,MN}$, then its inverse $\mathbf{X}^{-1} \in \mathcal{C}_{t^2,MN}$, and it always exists.
\end{lemma}
To prove \emph{Lemma} \ref{Lemma2}, we exploit i) the property from \eqref{eq:D_A} that the matrix $\mathbf{X}$ can be partitioned in terms of the diagonal blocks $\mathbf{X}_{i,j}$ for $1\leq i,j\leq t$, since $\mathbf{X}$ belongs to the set $\mathcal{C}_{t^2,MN}$; and ii) \emph{Lemma} \ref{Lemma1} that the \emph{Schur} complement of the diagonal matrix $\mathbf{X}_{t,t}$ belongs to the set $\mathcal{C}_{(t-1)^2,MN}$, and therefore, inverse of the \emph{Schur} complement can also be calculated using the block-wise inverse property from \eqref{eq:mimo_otfs_mtx1}. 

We now use \emph{Lemma} \ref{Lemma1}, \emph{Lemma} \ref{Lemma2} and the results given in \eqref{eq:D_A} and \eqref{eq:mimo_otfs_mtx1} to design a low-complexity Algorithm-\ref{algo:DA_INV} for computing the inverse $\mathbf{D}^{-1}_{\text{A}}$, i.e.,  $\mathbf{D}_{\text{LZ}}^{-1}$ and $\mathbf{D}_{\text{LM}}^{-1}$, for the proposed LM and LZ receivers. The  proposed Algorithm-\ref{algo:DA_INV} operates in two steps: a) matrix partitioning; and b) backtracking. We first explain the key ideas: 
\begin{itemize}
\item \textit{Matrix partitioning (lines $4-7$):} The matrix $\mathbf{D}_{\text{A}}$ is initially partitioned using \eqref{eq:D_A}. Subsequently, for $i=2,3,\ldots, N_t-1$, we exploit \emph{Lemma} \ref{Lemma1} and \eqref{eq:D_A} to partition the \emph{Schur} complement of $\mathbf{D}_{N_t-i}$. These partitions are stored in the corresponding matrices $\mathbf{P}_{N_t-i}$.
\item \textit{Backtracking phase (lines $9-13$):} Using the matrix partitions obtained from the first phase, the backtracking phase, as the name suggests, recursively computes the inverse of the matrix $\mathbf{D}_{\text{A}}$ by exploiting \emph{Lemma} \ref{Lemma2} and the block-wise inverse property from \eqref{eq:mimo_otfs_mtx1}.
\end{itemize}
We next explain the detailed operation of the proposed Algorithm-\ref{algo:DA_INV}.

 %Algorithm-\ref{algo:DA_INV} provides the pseudo-code for the computationally efficient calculation of .

\begin{algorithm}[htbp]
	\DontPrintSemicolon % Some LaTeX compilers require you to use \dontprintsemicolon instead 
	\KwIn{ Matrix $\mathbf{D}\in\mathcal{C}_{N_rN_t,MN}$}
	\KwOut{Matrix $\mathbf{D}_{\text{A}}^{-1}\in\mathcal{C}_{N^2_t,MN}$}
	Compute $\mathbf{D}_{\text{A}}=\mathbf{D}^H\mathbf{D}$, if $\text{A}\in\text{LZ}$; otherwise $\mathbf{D}_{\text{A}}=\mathbf{D}_{\text{LM}}=\mathbf{D}^H\mathbf{D}+\rho\mathbf{I}_{N_tMN}$\;
	\textbf{Initialization:} $\mathbf{P}_{N_t-1}=[\,\,],\mathbf{P}_{N_t-2}=[\,\,],\ldots,\mathbf{P}_{1}=[\,\,]$ and $\mathbf{P}_{\text{0}}=\mathbf{D}_{\text{A}}\in\mathcal{C}_{t^2,MN}$\;
	*/ Matrix Partitioning */\;
	\For{$i=1:1:N_t-1$} {
		Partition $\mathbf{P}_{0}$ using \eqref{eq:D_A} and store in $\mathbf{P}_{N_t-i}$ as 
		$
		\mathbf{P}_{N_t-i}=\begin{bmatrix}\mathbf{A}_{N_t-i}& \mathbf{B}_{N_t-i}\\ 
		\mathbf{C}_{N_t-i}& \mathbf{D}_{N_t-i} 
		\end{bmatrix}.
		$\;
		Replace $\mathbf{P}_{0}$ by the \emph{Schur} complement of $\mathbf{D}_{N_t-i}$ as $\mathbf{P}_{0}=\mathbf{A}_{N_t-i}-\mathbf{B}_{N_t-i}\mathbf{D}^{-1}_{N_t-i}\mathbf{C}_{N_t-i}$. \;
	}
	*/ Backtracking */  \;
	Compute $\mathbf{S}^{-1}_{1}=\mathbf{P}_0^{-1}=(\mathbf{A}_{1}-\mathbf{B}_{1}\mathbf{D}^{-1}_{1}\mathbf{C}_{1})^{-1}$\;
	\textbf{Initialize} the matrix $\mathbf{F}$  using \eqref{eq:mimo_otfs_mtx1} as   
	$\mathbf{F}=\begin{bmatrix}\mathbf{S}^{-1}_{1}& -\mathbf{S}^{-1}_{1}\mathbf{B}_1\mathbf{D}_1^{-1}\\ 
	-\mathbf{D}_1^{-1}\mathbf{C}_1\mathbf{S}^{-1}_{1}& \mathbf{D}_1^{-1}+\mathbf{D}_1^{-1}\mathbf{C}_1\mathbf{S}^{-1}_{1}\mathbf{B}_1\mathbf{D}_1^{-1} 
	\end{bmatrix}.$\;
	\For{$i=2:1:N_t-1$} {
		Compute block-wise inverse using \eqref{eq:mimo_otfs_mtx1} as
		$ \mathbf{S}_{i+1}^{-1}=\begin{bmatrix}\mathbf{F}& -\mathbf{F}\mathbf{B}_i\mathbf{D}_i^{-1}\\ 
		-\mathbf{D}_i^{-1}\mathbf{C}_i\mathbf{F}& \mathbf{D}_i^{-1}+\mathbf{D}_i^{-1}\mathbf{C}_i\mathbf{F}\mathbf{B}_i\mathbf{D}_i^{-1} 
		\end{bmatrix}$.\;
		Replace $\mathbf{F}= \mathbf{S}_{i+1}^{-1}$  \;
	}
	\Return:{\ $ \mathbf{D}_{\text{A}}^{-1}=\mathbf{F}$}\;
	\caption{{\sc } 	\footnotesize
		Pseudo-code for computing the inverse of matrix $\mathbf{D}_{\text{A}}\in\mathcal{C}_{N_t^2,MN}$, where $\text{A}\in\{\text{LZ, LM}\}$}
	\label{algo:DA_INV}
\end{algorithm}

\paragraph{Matrix Partitioning (lines $4-7$)} It  is an iterative process. Initially (for $i=1$), by using \eqref{eq:D_A}, we partition $\mathbf{D}_{\text{A}}\in\mathcal{C}_{N_t^2,MN}$ in terms of the sub-matrices $\mathbf{A}_{N_t-i}\in\mathbb{C}^{(N_t-i)MN\times (N_t-i)MN}$, $\mathbf{B}_{N_t-i}\in\mathbb{C}^{(N_t-i)MN\times MN}$, $\mathbf{C}_{N_t-i}\in\mathbb{C}^{MN\times (N_t-i)MN}$ and $\mathbf{D}_{N_t-i}\in\mathbb{C}^{MN\times MN}$, and store these partitions in the matrix  $\mathbf{P}_{N_t-1}$. For the block-wise inversion of $\mathbf{D}_{\text{A}}$ according to \eqref{eq:mimo_otfs_mtx1}, we now need  to invert  the \emph{Schur} complement $\mathbf{S}_{N_t-1}=(\mathbf{A}_{N_t-1}-\mathbf{B}_{N_t-1}\mathbf{D}^{-1}_{N_t-1}\mathbf{C}_{N_t-1})$. We know from \emph{Lemma} \ref{Lemma1} that $\mathbf{S}_{N_t-1}$ belongs to the set $\mathcal{C}_{(N_t-1)^2,MN}$. It can therefore be inverted block-wise using \eqref{eq:D_A} and \eqref{eq:mimo_otfs_mtx1}.  We, thus, partition $\mathbf{S}_{N_t-1}$ in terms of $\mathbf{A}_{N_t-2}$, $\mathbf{B}_{N_t-2}$, $\mathbf{C}_{N_t-2}$ and $\mathbf{D}_{N_t-2}$ using \eqref{eq:D_A}, and subsequently evaluate $\mathbf{S}_{N_t-2}=(\mathbf{A}_{N_t-2}-\mathbf{B}_{N_t-2}\mathbf{D}^{-1}_{N_t-2}\mathbf{C}_{N_t-2})$, which, from \emph{Lemma} \ref{Lemma1}, belongs to the set $ \mathcal{C}_{(N_t-2)^2,MN}$. We, therefore, next split $\mathbf{S}_{N_t-2}$ in terms of the sub-matrices $\mathbf{A}_{N_t-3}$, $\mathbf{B}_{N_t-3}$, $\mathbf{C}_{N_t-3}$ and $\mathbf{D}_{N_t-3}$ for computing $\mathbf{S}_{N_t-3}$ and so on. To summarize, for $i=2,\ldots,N_t-1$, we partition the \emph{Schur} complement $\mathbf{S}_{N_t-i}$ using \eqref{eq:D_A}, and store these partitions in the corresponding matrices $\mathbf{P}_{N_t-i}$. With $i=N_t-1$, we reach $\mathbf{S}_1$ which is an $MN\times MN$ diagonal matrix, and therefore, $\mathbf{S}^{-1}_1$ does not require further partitioning.

\paragraph{Backtracking (lines $9-13$)}  The partitioned matrix obtained in the first phase is now used to calculate $\mathbf{D}_{\text{A}}^{-1}$ by exploiting \emph{Lemma} \ref{Lemma2} and the result from \eqref{eq:mimo_otfs_mtx1}. As shown in the line-$10$, we initialize backtracking process by the matrix $\mathbf{F}$, which is obtained using $\mathbf{S}^{-1}_1$ and the sub-matrices $\{\mathbf{A}_1,\mathbf{B}_1,\mathbf{C}_1,\mathbf{D}_1\}$ stored in the matrix $\mathbf{P}_1$. Note that $\mathbf{F}$ is  the inverse of the \emph{Schur} complement $\mathbf{S}_2$. We know from \emph{Lemma} \ref{Lemma2} that if $\mathbf{S}_i$ belongs to the set $\mathcal{C}_{i^2,MN}$, its inverse $\mathbf{S}_i^{-1}$ also belongs to the same set $\mathcal{C}_{i^2,MN}$. Exploiting this property and the block-wise inverse from \eqref{eq:mimo_otfs_mtx1}, in the lines $11-13$, we recursively compute $\mathbf{S}_{i+1}^{-1}$ using  $\mathbf{S}_{i}^{-1}$ and the sub-matrices $\{\mathbf{A}_i,\mathbf{B}_i,\mathbf{C}_i,\mathbf{D}_i\}$  stored in $\mathbf{P}_i$. We see  that, when $i=N_t-1$, the inverse $\mathbf{S}^{-1}_{N_t}$ in the line $12$ is computed with the help of  \emph{Schur} complement $\mathbf{F}=\mathbf{S}_{N_t-1}^{-1}$ and the sub-matrices $\{\mathbf{A}_{N_t-1},\mathbf{B}_{N_t-1},\mathbf{C}_{N_t-1},\mathbf{D}_{N_t-1}\}$~as 
\begin{align}\label{eq:S_Nt}
\mathbf{S^{-1}_{N_t}}=\begin{bmatrix}\mathbf{S}_{N_t-1}^{-1}& -\mathbf{S}_{N_t-1}^{-1}\mathbf{B}_{N_t-1}\mathbf{D}_{N_t-1}^{-1}\\ 
-\mathbf{D}_{N_t-1}^{-1}\mathbf{C}_{N_t-1}\mathbf{S}_{N_t-1}^{-1}& \mathbf{D}_{N_t-1}^{-1}+\mathbf{D}_{N_t-1}^{-1}\mathbf{C}_{N_t-1}\mathbf{S}_{N_t-1}^{-1}\mathbf{B}_{N_t-1}\mathbf{D}_{N_t-1}^{-1} 
\end{bmatrix}.
\end{align}
 By comparing \eqref{eq:S_Nt} with \eqref{eq:mimo_otfs_mtx1}, we see that $\mathbf{S^{-1}_{N_t}}= \mathbf{D}_{\text{A}}^{-1}$. In other words, in the final step of the backtracking phase, the matrix $\mathbf{F}$ returns $\mathbf{D}^{-1}_{\text{A}}$. 
%\begin{observation}
%\colr{Sir, we do not need this observation when we are explaining this again in the next section after proof of Lemma 4.}As shown in the backtracking phase of the proposed Algorithm-\ref{algo:DA_INV},  computation of $\mathbf{D}_{\text{A}}^{-1}$ requires i) inversion of $MN\times MN$ diagonal matrices $\mathbf{D}_i$, which can be performed $MN$ multiplications only; and ii) multiplication of blocks matrices $\{\mathbf{F},\mathbf{A}_i,\mathbf{B}_i,\mathbf{C}_i,\mathbf{D}_i\}$, whose each block is an $MN\times MN$ diagonal matrix. This clearly shows the computational efficiency of the proposed algorithm.
%\end{observation}
{\emph{The proposed LZ and LM receivers thus give exactly the same solution as that the conventional ZF and MMSE receivers. Their complexity, as shown next , due to Algorithm \ref{algo:DA_INV}, which computes the inverse, is significantly lower than their conventional counterparts.}} %\colb{Given the observation 3, I thank we do not need this statement now}

\section{Complexity of the proposed low-complexity ZF and MMSE receivers}\label{complexity}
We compute the computational complexity by considering multiplication/division and addition/subtraction as  operations \cite{SurabhiC20}. {Note that the operations of the proposed receivers remain same regardless of perfect/imperfect CSIs, so is the computational complexity.} 
\subsubsection{Computation of $\mathbf{D}_{\text{A}}^{-1}$}\label{complexity1} For evaluating the complexity of the operation $\mathbf{D}_{\text{A}}^{-1}$, we state the following lemma, which is proved in Appendix \ref{Lemma2_1_proof}.
\begin{lemma}\label{Lemma2_1}
Total number of operations required for computing $\mathbf{D}_{\text{A}}^{-1}$, for $\text{A}\in\text{\{LZ, LM\}}$, is 
 \begin{align}
 \nonumber \mu_{\text{D}_{\text{LZ}}} &=[2N_t^3-3N_t^2+N_t+2N_t^2N_r]MN\ \text{ and }\ \mu_{\text{D}_{\text{LM}}} =[2N_t^3-3N_t^2+3N_t+2N_t^2N_r]MN.
 \end{align}
\end{lemma}
We see that the computational complexity for calculating $\mathbf{D}_{\text{A}}^{-1}$ with Algorithm-\ref{algo:DA_INV} is $\mathcal{O}(MN)$, since in practice $N_t,N_r \ll MN $. This is unlike conventional ZF and MMSE receivers which calculate their respective inverse operations with $\mathcal{O}(N_t^3M^3N^3)$ complexity.  We now evaluate the complexity of the proposed LZ and LM receivers in \eqref{eq:Combiners1} for processing the received vector $\mathbf{y}$, i.e., the operation $\mathbf{G}_{\text{A}}^{H}\mathbf{y}$  for $\text{A}\in\text{\{LZ, LM\}}$.
\subsubsection{Computation of $\mathbf{G}_{\text{A}}^{H}\mathbf{y}$}\label{complexity2}
%We evaluate the complexity for the ZF and MMSE processing of the observation vector $\mathbf{y}$ using
We state the following lemma which is proved in Appendix~\ref{Lemma2_2_proof}.
\begin{lemma}\label{Lemma2_2}
Total number of operations required for $\mathbf{G}_{\text{A}}^{H}\mathbf{y}$  operation is given as follows
\begin{align}
\mu_{\text{G}_{\text{A}}} = \big[2N_t^2N_r+N_tN_r-N_t\big]MN+\big[N_t+N_r\big]\mathcal{O}(MN\mbox{log}_2MN).
\end{align}
\end{lemma}
{Also i) calculation of the matrix $\mathbf{D}_{r,t}^i$ in the expression $\mathbf{D}_{r,t}=\sum_{i=1}^M\mathbf{\Omega}_M^i\otimes \mathbf{D}_{r,t}^i$ requires $\mathcal{O}(MN\log_2N)$ operations \cite{SurabhiAC19}; and ii) $\mathcal{O}(MN\log_2M)$ operations are needed to compute the eigenvalue matrix $\mathbf{D}_{r,t}$ of the matrix $\mathbf{H}_{r,t}$ \cite{SurabhiAC19}.
The complexity of computing the eigenvalue matrix $\mathbf{D}$ for the MIMO-OTFS channel matrix $\mathbf{H}$ is thus $N_rN_t\mathcal{O}(MN\log_2MN)$.} By exploiting this result, \emph{Lemma} \ref{Lemma2_1} and \emph{Lemma} \ref{Lemma2_2},  we calculate the overall receiver complexity~as %computational load for the proposed LC-ZF and LC-MMSE receivers for MIMO-OTFS as
\begin{align}
\mu_{\text{LZ}} = \mu_{\text{D}_{\text{LZ}}}+\mu_{\text{G}_{\text{A}}}&=\big[2N_t^3-3N_t^2+4N_t^2N_r+N_tN_r\big]MN\nonumber \\
& +\big[N_t+N_r+N_tN_r\big]\mathcal{O}(MN\mbox{log}_2MN)\nonumber \\
 \mu_{\text{LM}} &= \mu_{\text{LZ}}+2N_tMN.\label{eq:overall_complexity_MMSE}
\end{align}
We observe from  \eqref{eq:overall_complexity_MMSE} that the complexity of  LZ and LM receivers {is significantly lower than the conventional ZF and MMSE receivers, which have $\mathcal{O}(N_t^3M^3N^3)$ complexity. Further, the complexity of the MP-based receiver in SISO-OTFS systems varies as $\mathcal{O}(N_IMNSQ)$ \cite{raviteja2018interference}, where $N_I$ is  the number iterations required for the MP algorithm to converge, $Q$ is the constellation size, and $S$ is the number of non-zero elements in each row or column of SISO-OTFS channel matrix $\mathbf{H}_{\text{s}}\in\mathbb{C}^{MN\times MN}$. Since there are $N_r N_t$ number of links between the MIMO-OTFS transmitter and receiver, the complexity of the MP algorithm for data detection in MIMO-OTFS systems can be obtained as  $\mathcal{O}(N_IN_rN_tMNSQ)$, which is significantly higher than the proposed LZ and LM designs, as also shown numerically in Section-\ref{Results}. %The computational complexities of the proposed low-complexity and the conventional receivers is numerically also  compared in Section-\ref{Results}.
\section{BER derivation for LZ and LM receivers with imperfect receive CSI}\label{BER}
%We now derive the BER for the ZF and MMSE receivers for MIMO-OTFS for imperfect receive CSI. 
Recall that the SINR expressions of the conventional ZF and MMSE receivers for MIMO-OTFS  systems will   invert $\mathbf{H}^H\mathbf{H}$ and $\mathbf{H}^H\mathbf{H}+\rho \mathbf{I}$ matrices respectively, with  an extremely high computational complexity of $\mathcal{O}(N_t^3M^3N^3)$ \cite{SurabhiC20}. We now propose a low-complexity method for calculating the SINR of the proposed  LM and LZ receivers  with imperfect CSI using Algorithm~\ref{algo:DA_INV} .
We achieve this goal {by first deriving a tight approximation of their SINR expressions in closed-form}  and later showing that the Algorithm \ref{algo:DA_INV} can be used to calculate them. We begin by modeling the estimate of MIMO-OTFS channel  $\mathbf{H}$, denoted as $ \widehat{\mathbf{H}}$ \cite{WangAMMCL07},  as follows 
\begin{eqnarray}\label{eq:CFR_EST}
 \widehat{\mathbf{H}}= \mathbf{H}+\Delta\mathbf{H}.
\end{eqnarray}
The matrix $\Delta\mathbf{H}\in \mathbb{C}^{N_{r}MN\times N_{t}MN}$ is  the CSI error, which is independent of $\mathbf{H}$ \cite{WangAMMCL07}.  Since $\mathbf{H}$ belongs to the set $\mathcal{C}_{N_rN_t,MN}$, $\Delta\mathbf{H}$ also belongs to the set $\mathcal{C}_{N_rN_t,MN}$.  It follows from \emph{Lemma} \ref{Lemma1} that $\widehat{\mathbf{H}}$ also belongs to the set $\mathcal{C}_{N_rN_t,MN}$. Let $\Delta\mathbf{H}_{r,t}\in \mathbb{C}^{MN\times MN}$ be the error matrix corresponding to the channel $\mathbf{H}_{r,t}$ between the $t$th transmit and $r$th receive antennas. Since $\Delta\mathbf{H}_{r,t}$ belongs to the set $\mathcal{B}_{M,N}$, it consists of $M$ circulant blocks, each of size $N\times N$. We assume that the non-zero entries in a row or column of the block $\Delta\mathbf{H}_{r,t}$ of channel matrix $\Delta\mathbf{H}$ are  i.i.d. as $\mathcal{CN}(0,\sigma^{2}_{e})$ \cite{WangAMMCL07}. The variance $\sigma^{2}_{e}$ captures the accuracy of a channel estimator. The error matrix $\Delta\mathbf{H}$, similar to the decomposition of $\mathbf{H}$ in \eqref{eq:mimo_otfs_mtx_D},  can be decomposed as
\begin{equation}\label{eq:mimo_otfs_mtx_DeltaD}
\Delta\mathbf{H} = \boldsymbol{\Psi}_{\text{R}}^H\Delta\mathbf{D}\boldsymbol{\Psi}_{\text{T}},
\end{equation}
where block matrix $\Delta\mathbf{D}\in\mathbb{C}^{N_rMN\times N_tMN}$ contains eigenvalues of the error matrix $\Delta\mathbf{H}$. {Using  \eqref{eq:mimo_otfs_mtx_D} and \eqref{eq:mimo_otfs_mtx_DeltaD}, the MIMO-OTFS channel estimate $\widehat{\mathbf{H}}$ in  \eqref{eq:CFR_EST} can be re-written~as
\begin{equation}\label{eq:Decom_Dhat}
\widehat{\mathbf{H}} = \boldsymbol{\Psi}_{\text{R}}^H\widehat{\mathbf{D}}\boldsymbol{\Psi}_{\text{T}}.
\end{equation}
Here $\widehat{\mathbf{D}}=\mathbf{D}+\Delta\mathbf{D}$ is the estimate of $\mathbf{D}$.} We next derive the SINR of LZ and LM receivers.

\subsection{SINR calculation for the proposed LM receiver with imperfect receive CSI}\label{MMSE_SINR}
We first {tightly approximate the SINR expression in closed-form} for the proposed LM receiver in terms of interference, and noise-plus-distortion due to channel estimation error. We then derive a tractable SINR expression for the proposed LM receiver by exploiting the Taylor expansion, decompositions in \eqref{eq:mimo_otfs_mtx_D} and the properties of the sets defined in Table-\ref{table:sets}. We begin by substituting \eqref{eq:Decom_Dhat} in \eqref{eq:Combiners1} to express the proposed LM receiver with CSI error as follows
\begin{align}\label{eq:MMSE_COMB_ALT_EST}
 \widehat{\mathbf{G}}_{\text{LM}}=  \boldsymbol{\Psi}_{\text{R}}^H\widehat{\mathbf{D}}\big( \widehat{\mathbf{D}}^{H} \widehat{\mathbf{D}}+\rho\mathbf{I}_{N_{t}MN}\big)^{-1}\boldsymbol{\Psi}_{\text{T}}
 =\mathbf{G}_{\text{LM}}+\Delta\mathbf{G}_{\text{LM}}.
\end{align}
The $\Delta\mathbf{G}_{\text{LM}}$ expression is calculated later in the sequel.  The LM estimate of the transmit vector $\mathbf{x}$ with CSI error can now be obtained using \eqref{eq:mimo_otfs} as follows
\begin{align}
\hat{\mathbf{x}}_{\text{LM}}&= \widehat{\mathbf{G}}_{\text{LM}}^{H}\mathbf{y}= \mathbf{G}_{\text{LM}}^{H}\mathbf{H}\mathbf{x}+\mathbf{v}.\notag
\end{align}
Here $\mathbf{v}=\mathbf{G}^{H}_{\text{LM}}\tilde{\mathbf{v}}+\Delta\mathbf{G}^{H}_{\text{LM}}\mathbf{H}\mathbf{x}+\Delta\mathbf{G}^{H}_{\text{LM}}\tilde{\mathbf{v}}$ is the noise-plus-distortion due to  channel estimation errors. The estimated vector $\hat{\mathbf{x}}_{\text{LM}}\in \mathbb{C}^{N_tMN\times 1}$ consists of $N_tMN$ symbols. Let $\hat{x}^{m_{tk}}_{\text{LM}}$ denotes the estimate of $k$-th symbol of the $t$-th stream (antenna), where $m_{tk}=MN(t-1)+k$ with $1\leq k\leq MN$ and $1\leq t\leq N_t$. It implies that $1\leq m_{tk}\leq N_tMN$. The $\hat{x}^{m_{tk}}_{\text{LM}}$ is therefore 
\begin{align}
\nonumber \hat{x}^{m_{tk}}_{\text{LM}}= [\mathbf{G}_{\text{LM}}^{H}\mathbf{H}]_{(m_{tk},m_{tk})}{x}^{m_{tk}}+\underbrace{\sum_{i=(t-1)MN+1, i\neq k}^{tMN}[\mathbf{G}_{\text{LM}}^{H}\mathbf{H}]_{(m_{tk},i)}{x}^{i}}_{\text{inter-symbol interference}}+\underbrace{\sum_{j=1,j\neq k}^{N_tMN}[\mathbf{G}_{\text{LM}}^{H}\mathbf{H}]_{(m_{tk},j)}{x}^{j}+v^{m_{tk}}}_{\text{inter-stream interference}}.
\end{align}
We see that the estimated symbol $\hat{x}^{m_{tk}}_{\text{LM}}$ experiences inter-stream interference from $(N_t-1)MN$ symbols of other antennas \cite{SinghMJV19} and inter-symbol (intra-stream) interference from $MN-1$ symbols of the same antenna \cite{raviteja2018interference}. {The  inter-symbol interference occurs because of non-zero delay and Doppler spreads~\cite{raviteja2018interference}. The SINR of the $k$th symbol of $t$th stream, by using the property $\mathbb{E}[\mathbf{x}\mathbf{x}^H]=P_x\mathbf{I}_{N_t MN}$, can be expressed as
 \begin{align*}
\gamma^{m_{tk}}_{\text{LM}}=\dfrac{P_{x}\left|\left[\mathbf{G}_{\text{LM}}^{H}\mathbf{H}\right]_{(m_{tk},m_{tk})}\right|^2}{\displaystyle P_x\sum_{i=(t-1)MN+1,i\neq k}^{tMN}\left|\left[\mathbf{G}_{\text{LM}}^{H}\mathbf{H}\right]_{(m_{tk},i)}\right|^2+ P_{x}\sum_{j=1,j\neq i}^{N_tMN}\left|\left[\mathbf{G}_{\text{LM}}^{H}\mathbf{H}\right]_{(m_{tk},j)}\right|^2+\big[\mathbf{R}_{v}\big]_{(m_{tk},m_{tk})}}.
\end{align*}
{By substituting the expressions of  $\mathbf{G}_{\text{LM}}=\boldsymbol{\Psi}^H_\text{R}\mathbf{D}\left(\mathbf{D}^H\mathbf{D}+\rho\mathbf{I}_{N_tMN}\right)^{-1}\boldsymbol{\Psi}_\text{T}$ from \eqref{eq:Combiners1} and $\mathbf{H}$ from \eqref{eq:mimo_otfs_mtx_D}, the above SINR expression can be re-expressed as follows
\begin{footnotesize}
 \begin{align}\label{eq:PP-SINR_ICSI_fianl}
\gamma^{m_{tk}}_{\text{LM}}=\dfrac{P_{x}\left|\left[\boldsymbol{\Psi}_{\text{T}}^{H}\mathbf{D}_{\text{LM}}^{-1}\mathbf{D}_{\text{lcZF}}\boldsymbol{\Psi}_{\text{T}}\right]_{(m_{tk},m_{tk})}\right|^2}{\displaystyle P_x\sum_{i=(t-1)MN+1,i\neq k}^{tMN}\left|\left[\boldsymbol{\Psi}_{\text{T}}^{H}\mathbf{D}_{\text{LM}}^{-1}\mathbf{D}_{\text{lcZF}}\boldsymbol{\Psi}_{\text{T}}\right]_{(m_{tk},i)}\right|^2+ P_{x}\sum_{j=1,j\neq i}^{N_tMN}\left|\left[\boldsymbol{\Psi}_{\text{T}}^{H}\mathbf{D}_{\text{LM}}^{-1}\mathbf{D}_{\text{lcZF}}\boldsymbol{\Psi}_{\text{T}}\right]_{(m_{tk},j)}\right|^2+\big[\mathbf{R}_{v}\big]_{(m_{tk},m_{tk})}}.
\end{align}
\end{footnotesize}The first and second terms in the denominator of $\gamma^{m_{tk}}_{\text{LM}}$ denote the power of inter-symbol  and inter-stream interferences, respectively. We note that $\mathbf{D}_{\text{LM}}=\mathbf{D}^H\mathbf{D}+\rho\mathbf{I}_{N_t MN}$ and $\mathbf{D}_{\text{LZ}}=\mathbf{D}^H\mathbf{D}$.} 
\begin{remark}
Recall that Algorithm~\ref{algo:DA_INV} calculates low-complexity inverse of $\mathbf{D}_{\text{LM}}$ with $\mathcal{O}(MN)$ complexity. To use it, we need to show that the SINR expressions invert $\mathbf{D}_{\text{LM}}$ matrix. We see from \eqref{eq:PP-SINR_ICSI_fianl} that all the terms of LM SINR expression, except the noise-plus-interference covariance matrix $\mathbf{R}_v$, invert  $\mathbf{D}_{\text{LM}}$ matrices. We next simplify $\mathbf{R}_v$  using results from random matrix theory, and show that the simplified expression inverts $\mathbf{D}_{\text{LM}}$ matrix. This  enables us to apply Algorithm \ref{algo:DA_INV} for calculating SINR.  
\end{remark}
For a fixed channel realization, the covariance matrix $\mathbf{R}_{v}=\mathbb{E}\big[\mathbf{v}\mathbf{v}^{H}\big]$ can be evaluated as follows
\begin{equation}\label{eq:N_P_D_COV}
 \mathbf{R}_{v}= P_{x}\mathbb{E}\big[\Delta\mathbf{G}^{H}_{\text{LM}}\mathbf{H}\mathbf{H}^{H}\Delta\mathbf{G}_{\text{LM}}\big]+\sigma^{2}_{v}\mathbf{G}^{H}_{\text{LM}}\mathbf{G}_{\text{LM}}+\sigma^{2}_{v}\mathbb{E}\big[\Delta\mathbf{G}^{H}_{\text{LM}}\Delta\mathbf{G}_{\text{LM}}\big].
\end{equation}
Since $\mathbf{R}_{v}$ is a function of the matrix $\Delta\mathbf{G}_{\text{LM}}$, we first calculate the $\Delta\mathbf{G}_{\text{LM}}$ expression, and then use it for simplifying both first and third terms in \eqref{eq:N_P_D_COV}. 

\begin{figure}[htbp]
	\begin{center}
	\subfloat[]{\includegraphics[scale = 0.75]{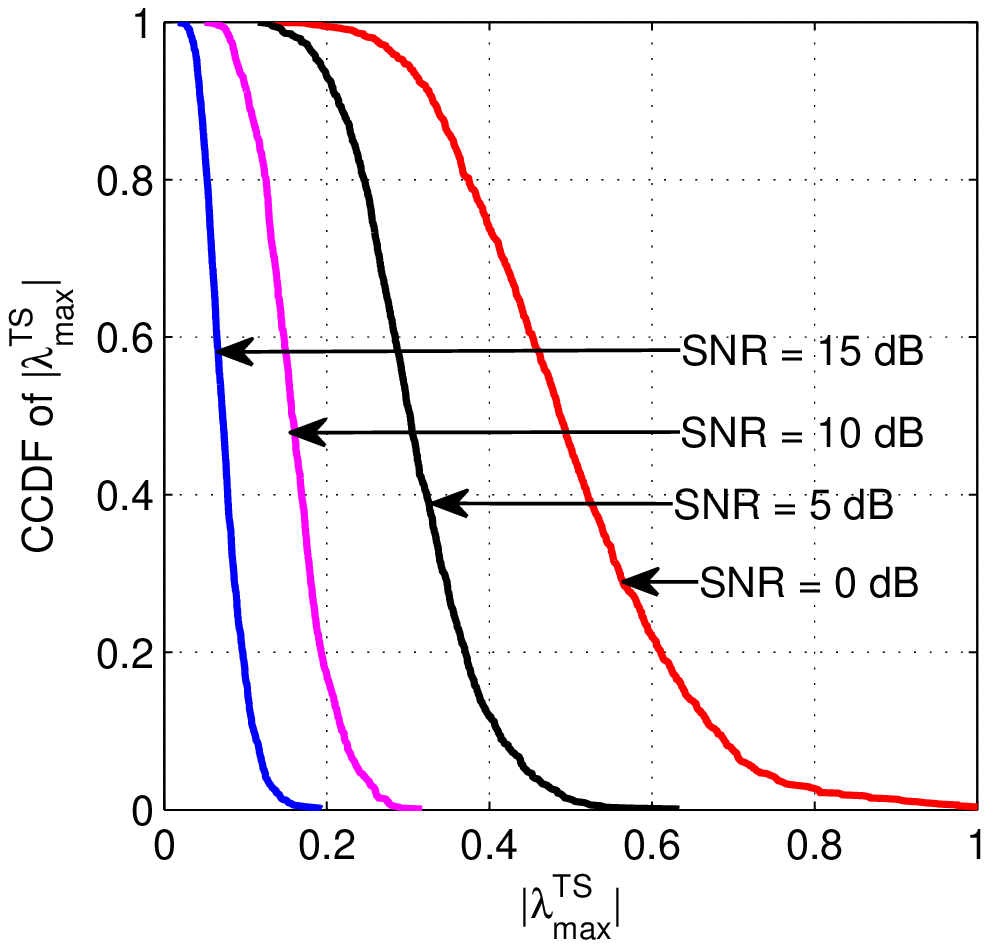}} 
	\hfil 
	\hspace{-10pt}\subfloat[]{\includegraphics[scale = 0.75]{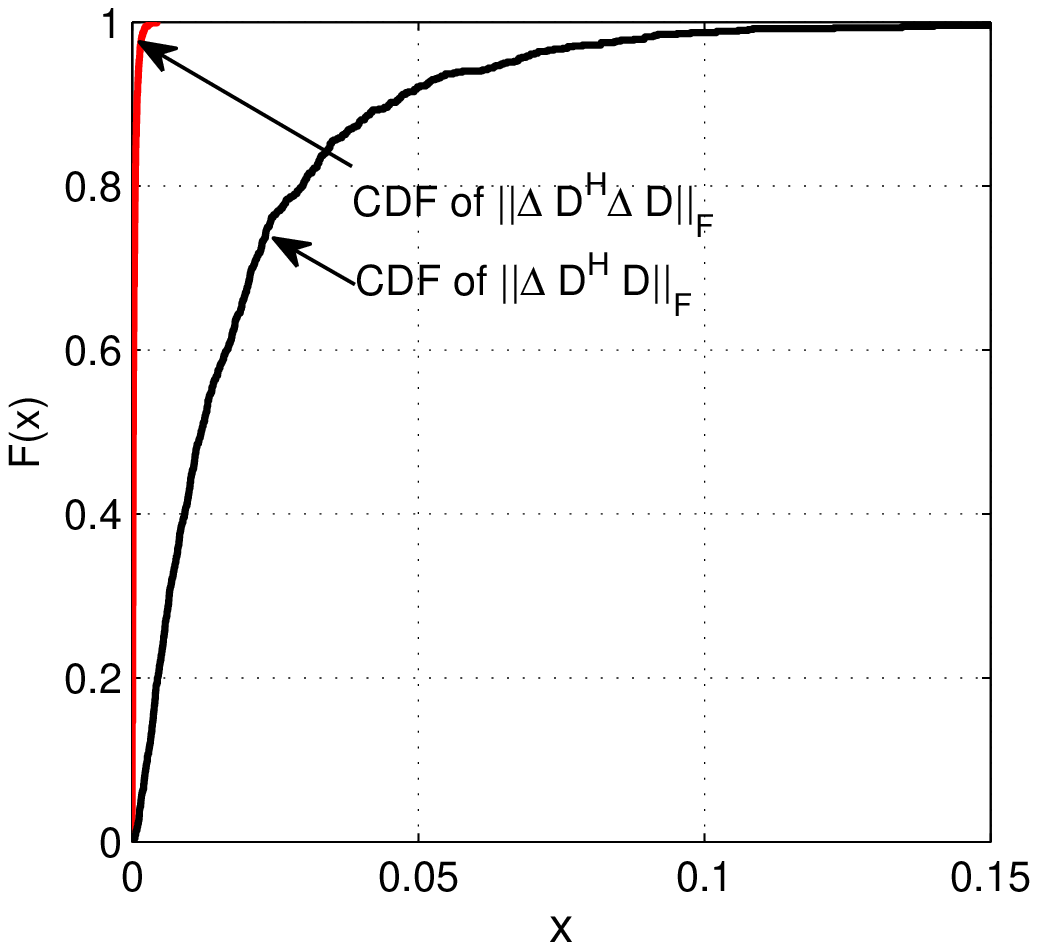}}
	\end{center}	
	\hfil
	\caption{\footnotesize{(a) Empirical CCDF of absolute value of maximum eigenvalue $\lambda_{\text{max}}^{\text{TS}}$ of  $\mathbf{TS}$ with $M=N=8$ and different SNR values; and (a) Empirical CDF of Frobenius norm of the matrix $\mathbf{\Delta D}^H\mathbf{\Delta D}$ and $\mathbf{\Delta D}^H\mathbf{D}$ for $M=N=8$ and SNR $=10$ dB.}}
	\label{fig:CDF}
\end{figure}
\subsubsection{{Expression for $\Delta\mathbf{G}_{\text{LM}}$}}
The matrix $\widehat{\mathbf{G}}^{H}_{\text{LM}}$, as shown in Appendix~\ref{proof_DeltaG},  is a function of the matrix $( \mathbf{S}^{-1}+ \mathbf{T})^{-1}$, where $\mathbf{S}^{-1}=\mathbf{D}_{\text{LM}}=\mathbf{D}^H\mathbf{D}+\rho\mathbf{I}_{N_tMN}$ and $\mathbf{T}=\Delta\mathbf{D}^{H} \mathbf{D}+\mathbf{D}^{H}\Delta\mathbf{D}$. The matrix  $\mathbf{\Delta D}$ in  $\mathbf{T}$ consists of eigenvalues of the error matrix $\mathbf{\Delta H}$.  Let $\lambda^{\text{TS}}_{\max}$ be the maximum eigenvalue of the matrix $\mathbf{TS}$. As shown in Fig.~\ref{fig:CDF}(a), the probability that $\lambda^{\text{TS}}_{\max}>1$  is close to zero.
The expression $( \mathbf{S}^{-1}+ \mathbf{T})^{-1}$ can thus be approximated  using its first order Taylor series expansion \cite{petersen2012matrix}. 
Also, as shown in Fig.~\ref{fig:CDF}(b), the probability that $||\mathbf{\Delta D}^H\mathbf{\Delta D}||_F\ll ||\mathbf{\Delta D}^H\mathbf{D}||_F$,  is close to one. The matrix $\mathbf{\Delta D}^H\mathbf{\Delta D}$ can thus be neglected. Fig.~\ref{fig:CDF}(a) and Fig.~1(b) also imply that our approximate SINR expression is tight.  The matrix $\mathbf{\Delta G}_{\text{LM}}$ expression, derived using above properties in Appendix~\ref{proof_DeltaG},  is given as follows
\begin{align}
\Delta\mathbf{G}^{H}_{\text{LM}}&\approx-\boldsymbol{\Psi}^{H}_{\text{T}}\big[\mathbf{S}\big(\Delta\mathbf{D}^{H} \mathbf{D}+\mathbf{D}^{H}\Delta\mathbf{D}\big)\mathbf{S}\mathbf{D}^{H}
-\mathbf{S}\Delta\mathbf{D}^{H}\big]\boldsymbol{\Psi}_{\text{R}}.\label{eq:N_P_D_COV_3}
\end{align}
 We next use the above expression for $\Delta\mathbf{G}_{\text{LM}}$ to calculate covariance matrix $\mathbf{R}_{v}$ in \eqref{eq:N_P_D_COV}.
\subsubsection{Expression for $\mathbf{R}_{v}$}
 The first and third terms of $\mathbf{R}_{v}$ in \eqref{eq:N_P_D_COV} are derived by substituting $\Delta\mathbf{G}_{\text{LM}}$ from \eqref{eq:N_P_D_COV_3}, and by using the decomposition from \eqref{eq:mimo_otfs_mtx_D} as
\begin{align}\label{eq:N_P_D_COV_4}
\nonumber  & \mathbb{E}\big[\Delta\mathbf{G}^{H}_{\text{LM}}\mathbf{D}\mathbf{D}^{H}\Delta\mathbf{G}_{\text{LM}}\big] \approx \boldsymbol{\Psi}_{\text{T}}^{H}\Big\{\mathbf{S}\mathbb{E}\big[\Delta\mathbf{D}^{H}\mathbf{D}_2\mathbf{D}_{1}\mathbf{D}_2^H\Delta\mathbf{D}\big]\mathbf{S}^{H}- \mathbf{S}\mathbb{E}\big[\Delta\mathbf{D}^{H}\mathbf{D}_{2}\mathbf{D}_{1}\Delta\mathbf{D}\big]\mathbf{S}^{H}\\
\nonumber &- \mathbf{D}^{H}_3\mathbb{E}\big[\Delta\mathbf{D}\mathbf{D}^{H}_3\mathbf{D}_{1}\Delta\mathbf{D}\big]\mathbf{S}^{H}+ \mathbf{S}\mathbb{E}\big[\Delta\mathbf{D}^{H}\mathbf{D}_1\Delta\mathbf{D}\big]\mathbf{S}^{H} +  \mathbf{D}_3^{H}\mathbb{E}\big[\Delta\mathbf{D}\mathbf{D}_{3}^H\mathbf{D}_{1}\mathbf{D}^H_{2}\Delta\mathbf{D}\big]\mathbf{S}^{H}\\
\nonumber &- \mathbf{S}\mathbb{E}\big[\Delta\mathbf{D}^{H}\mathbf{D}_{1}\mathbf{D}_{2}^{H}\Delta\mathbf{D}\big]\mathbf{S}^{H}+\mathbf{S}\mathbb{E}\big[\Delta\mathbf{D}^{H}\mathbf{D}_{2}\mathbf{D}_{1}\mathbf{D}_{3}\Delta\mathbf{D}^{H}\big]\mathbf{D}_{3} + \mathbf{D}_3^{H}\mathbb{E}\big[\Delta\mathbf{D}\mathbf{D}_{3}^{H}\mathbf{D}_{1}\mathbf{D}_3\Delta\mathbf{D}^{H}\big]\mathbf{D}_3 \\
&- \mathbf{S}\mathbb{E}\big[\Delta\mathbf{D}^{H}\mathbf{D}_{1}\mathbf{D}_{3}\Delta\mathbf{D}^{H}\big] \mathbf{D}_3 \Big\}\boldsymbol{\Psi}_{\text{T}}\ \ \text{and}\\
&\nonumber \mathbb{E}\big[\Delta\mathbf{G}^{H}_{\text{LM}}\Delta\mathbf{G}_{\text{LM}}\big] \approx \boldsymbol{\Psi}_{\text{T}}^{H}\Big\{\mathbf{S}\mathbb{E}\big[\Delta\mathbf{D}^{H}\mathbf{D}_2\mathbf{D}_2^H\Delta\mathbf{D}\big]\mathbf{S}^{H}- \mathbf{S}\mathbb{E}\big[\Delta\mathbf{D}^{H}\mathbf{D}_{2}\Delta\mathbf{D}\big]\mathbf{S}^{H}\\
\nonumber &-\mathbf{D}^{H}_3\mathbb{E}\big[\Delta\mathbf{D}\mathbf{D}^{H}_3\Delta\mathbf{D}\big]\mathbf{S}^{H} + \mathbf{S}\mathbb{E}\big[\Delta\mathbf{D}^{H}\Delta\mathbf{D}\big]\mathbf{S}^{H} +  \mathbf{D}_3^{H}\mathbb{E}\big[\Delta\mathbf{D}\mathbf{D}_{3}^H\mathbf{D}^H_{2}\Delta\mathbf{D}\big]\mathbf{S}^{H}\\
\nonumber &- \mathbf{S}\mathbb{E}\big[\Delta\mathbf{D}^{H}\mathbf{D}_{2}^{H}\Delta\mathbf{D}\big]\mathbf{S}^{H}+\mathbf{S}\mathbb{E}\big[\Delta\mathbf{D}^{H}\mathbf{D}_{2}\mathbf{D}_{3}\Delta\mathbf{D}^{H}\big]\mathbf{D}_{3}+ \mathbf{D}_3^{H}\mathbb{E}\big[\Delta\mathbf{D}\mathbf{D}_{3}^{H}\mathbf{D}_3\Delta\mathbf{D}^{H}\big]\mathbf{D}_3 \\
&- \mathbf{S}\mathbb{E}\big[\Delta\mathbf{D}^{H}\mathbf{D}_{3}\Delta\mathbf{D}^{H}\big] \mathbf{D}_3 \Big\}\boldsymbol{\Psi}_{\text{T}}.\label{eq:N_P_D_COV_4_1}
\end{align}
Here $\mathbf{D}_1\in\mathbb{C}^{N_rMN\times N_rMN}$, $\mathbf{D}_2\in\mathbb{C}^{N_rMN\times N_rMN}$ and $\mathbf{D}_3\in\mathbb{C}^{N_rMN\times N_tMN}$ are defined as  $\mathbf{D}_{1}=\mathbf{D} \mathbf{D}^{H}$, $\mathbf{D}_{2}=\mathbf{D}\mathbf{S} \mathbf{D}^{H}$ and $\mathbf{D}_{3}=\mathbf{D} \mathbf{S}^{H}$. For evaluating the expectation operations in \eqref{eq:N_P_D_COV_4} and \eqref{eq:N_P_D_COV_4_1}, we state the following lemmas.
\begin{lemma}\label{Lemma3}
Elements of all the $N_rN_t$ blocks of the matrix $\Delta\mathbf{D}\in\mathcal{C}_{N_rN_t,MN}$ are i.i.d. as $\mathcal{CN}(0,\sigma^2_d)$, where $\sigma^2_d=\sigma^2_e \sum_{i=1}^{M}L^{i}_h$ with $L^{i}_h$ being the number of non-zero elements in a row or column of the $i$th circulant block of the $(r,t)$th sub-matrix $\Delta\mathbf{H}_{r,t}$ of $\Delta\mathbf{H}$. %\colr{check if it equal to $\sigma^2_eL_h$}
\end{lemma}
\begin{IEEEproof}
Refer to Appendix~\ref{Lemma3_proof}.
\end{IEEEproof}
To prove \emph{Lemma} \ref{Lemma3}, we exploit the i) error matrix decomposition in \eqref{eq:mimo_otfs_mtx_DeltaD}; ii) result in \eqref{eq:Prop2} for deriving the eigenvalues for $(r,t)$th block $\Delta\mathbf{H}_{r,t}$ of the error matrix $\Delta\mathbf{H}$; and iii) {property of Kronecker product of a Gaussian distributed matrix and a deterministic matrix  \cite{petersen2012matrix}.}   
\begin{lemma}\label{Lemma4}
For a random matrix $\mathbf{X}\in\mathcal{C}_{N_rN_t,MN}$ and deterministic matrices $\mathbf{Y}\in\mathcal{C}_{N_r^2,MN}$ and $\mathbf{Z}\in\mathcal{C}_{N_t^2,MN}$, if elements in all the $N_rN_t$ blocks of $\mathbf{X}$ are i.i.d. with mean zero and variance $\sigma^2_x$,
\begin{align}\label{eq:Lemma4_results}
\mathbb{E}[\mathbf{X}^{H}\mathbf{Y}\mathbf{X}]=\sigma^2_x\bar{\mathbf{Y}}\ \text{ and}\ \ 
\mathbb{E}[\mathbf{X}\mathbf{Z}\mathbf{X}^{H}]=\sigma^2_x\bar{\mathbf{Z}}.
\end{align}
{Here $\bar{\mathbf{Y}}=\sum_{i=1}^{N_r}(\mathbf{I}_{N_t}\otimes \mathbf{Y}_{i,i})$ and $\bar{\mathbf{Z}}=\sum_{i=1}^{N_t}(\mathbf{I}_{N_r}\otimes \mathbf{Z}_{i,i})$ are block diagonal matrices of size $N_tMN\times N_tMN$ and $N_rMN\times N_rMN$ respectively}, where $\mathbf{Y}_{i,i}\in\mathbb{C}^{MN\times MN}$ and $\mathbf{Z}_{i,i}\in\mathbb{C}^{MN\times MN}$ are $i$th diagonal block of the matrices $\mathbf{Y}$ and $\mathbf{Z}$, respectively.
\end{lemma}
\begin{IEEEproof}
Refer to Appendix~\ref{Lemma4_proof}.
\end{IEEEproof}
\begin{lemma}\label{Lemma5}
For a random matrix $\mathbf{X}\in\mathcal{C}_{N_rN_t,MN}$ and deterministic matrices $\mathbf{A},\mathbf{B}\in\mathcal{C}_{N_rN_t,MN}$, if elements in all the $N_rN_t$ blocks of $\mathbf{X}$ are i.i.d. with mean zero and variance $\sigma^2_x$, then $\mathbb{E}[\mathbf{X}\mathbf{A}\mathbf{X}]=\mathbf{0}_{N_rMN\times N_tMN}$ and $\mathbb{E}[\mathbf{X}^H\mathbf{B}\mathbf{X}^{H}]=\mathbf{0}_{N_tMN\times N_rMN}$.
\end{lemma}
\begin{IEEEproof}
{Proof follows from \emph{Lemma}~\ref{Lemma1} and the properties of complex random matrices.}
\end{IEEEproof}
{For deriving the results in \emph{Lemma} \ref{Lemma4} and \emph{Lemma} \ref{Lemma5}, we use \emph{Lemma} \ref{Lemma1}, statistical characteristics of the matrix $\mathbf{X}$, and the properties of the sets $\mathcal{C}_{N_rN_t,MN}$,  $\mathcal{C}_{N_r^2,MN}$ and  $\mathcal{C}_{N_t^2,MN}$.}
\subsubsection{Simplification of \eqref{eq:N_P_D_COV_4} and \eqref{eq:N_P_D_COV_4_1}} We use the below observations to simplify \eqref{eq:N_P_D_COV_4} and~\eqref{eq:N_P_D_COV_4_1}.
{\begin{observation}\label{obsrv:5} 
Since the matrices $\mathbf{D}_1,\mathbf{D}_2\in\mathcal{C}_{N_r^2,MN}$ and $\mathbf{D}_3\in\mathcal{C}_{N_rN_t,MN}$, it follows from \emph{Lemma} \ref{Lemma1} that, in \eqref{eq:N_P_D_COV_4}, the equivalent matrix i) multiplied by $\Delta\mathbf{D}$ from both sides belongs to the set $\mathcal{C}_{N_rN_t,MN}$;
ii) multiplied by $\Delta\mathbf{D}^H$ from both sides also belongs to the set $\mathcal{C}_{N_rN_t,MN}$; iii) sandwiched between $\Delta\mathbf{D}$ and $\Delta\mathbf{D}^{H}$ belongs to the set $\mathcal{C}_{N_t^2,MN}$; and  
iv) sandwiched between $\Delta\mathbf{D}^H$ and $\Delta\mathbf{D}$ belongs to the set $\mathcal{C}_{N_r^2,MN}$. 
\end{observation}}
{Applying \emph{Lemma} \ref{Lemma3}, \emph{Lemma} \ref{Lemma4}, \emph{Lemma} \ref{Lemma5} and \emph{Observation}~\ref{obsrv:5}, as shown in Appendix~\ref{Simplify_Rv}, the expression for $\mathbb{E}\big[\Delta\mathbf{G}^{H}_{\text{LM}}\mathbf{D}\mathbf{D}^{H}\Delta\mathbf{G}_{\text{LM}}\big]$ and $\mathbb{E}\big[\Delta\mathbf{G}^{H}_{\text{LM}}\Delta\mathbf{G}_{\text{LM}}\big]$ can be obtained~as
\begin{align}\label{eq:N_P_D_COV_5}
&\mathbb{E}\big[\Delta\mathbf{G}^{H}_{\text{LM}}\mathbf{D}\mathbf{D}^{H}\Delta\mathbf{G}_{\text{LM}}\big]  =  \sigma^2_d\boldsymbol{\Psi}_{\text{T}}^{H}\Big\{\mathbf{S}(\bar{\mathbf{D}}_{1}- \bar{\mathbf{D}}_{2}+ \bar{\mathbf{D}}_4- \bar{\mathbf{D}}_{6})\mathbf{S}^{H} + \mathbf{D}_3^{H}\bar{\mathbf{D}}_8\mathbf{D}_3 \Big\}\boldsymbol{\Psi}_{\text{T}},\\
 &\mathbb{E}\big[\Delta\mathbf{G}^{H}_{\text{LM}}\Delta\mathbf{G}_{\text{LM}}\big]=\sigma^2_d\boldsymbol{\Psi}_{\text{T}}^{H}\Big\{\mathbf{S}\tilde{\mathbf{D}}_{1}\mathbf{S}^{H}- \mathbf{S}\tilde{\mathbf{D}}_{2}\mathbf{S}^{H}+ N_r\mathbf{S}\mathbf{S}^{H}- \mathbf{S}\tilde{\mathbf{D}}_{6}\mathbf{S}^{H} + \mathbf{D}_3^{H}\tilde{\mathbf{D}}_8\mathbf{D}_3 \Big\}\boldsymbol{\Psi}_{\text{T}}.\label{eq:N_P_D_COV_6}
\end{align}
Here the block diagonal matrices $\{\bar{\mathbf{D}}_{1},\bar{\mathbf{D}}_{2},\bar{\mathbf{D}}_{4},\bar{\mathbf{D}}_{6}\}\in\mathbb{C}^{N_tMN\times N_tMN}$, $\bar{\mathbf{D}}_{8}\in\mathbb{C}^{N_rMN\times N_rMN}$, $\{\tilde{\mathbf{D}}_{1},\tilde{\mathbf{D}}_{2},\tilde{\mathbf{D}}_{6}\}\in\mathbb{C}^{N_tMN\times N_tMN}$ and $\tilde{\mathbf{D}}_{8}\in\mathbb{C}^{N_rMN\times N_rMN}$ are defined in Appendix~\ref{Simplify_Rv}. Using \eqref{eq:Combiners1}, the proposed LM receiver matrix can also be written as $\mathbf{G}_{\text{LM}}^{H}=\boldsymbol{\Psi}_{\text{T}}^H\mathbf{S}\mathbf{D}^H\boldsymbol{\Psi}_{\text{R}}=\boldsymbol{\Psi}_{\text{T}}^H\mathbf{D}_3^H\boldsymbol{\Psi}_{\text{R}}$. Thus, the second term of the covariance matrix $\mathbf{R}_{v}$ in \eqref{eq:N_P_D_COV} can be evaluated as $\sigma^2_{v}\mathbf{G}_{\text{LM}}^{H}\mathbf{G}_{\text{LM}}=\sigma^2_{v}\boldsymbol{\Psi}_{\text{T}}^H\mathbf{D}_3^H\mathbf{D}_3\boldsymbol{\Psi}_{\text{T}}$. Substituting \eqref{eq:N_P_D_COV_5}, \eqref{eq:N_P_D_COV_6} and the above expression in \eqref{eq:N_P_D_COV}, we get
\begin{align}\label{eq:R_v_final}
\nonumber   \mathbf{R}_{v} &=  P_x\sigma^2_d\boldsymbol{\Psi}_{\text{T}}^{H}\Big\{\mathbf{S}(\bar{\mathbf{D}}_{1}+\rho\tilde{\mathbf{D}}_{1})\mathbf{S}^{H}- \mathbf{S}(\bar{\mathbf{D}}_{2}+\rho\tilde{\mathbf{D}}_{2})\mathbf{S}^{H}+ \mathbf{S}(\bar{\mathbf{D}}_4+\rho N_r\mathbf{I}_{N_tMN})\mathbf{S}^{H}\\
&- \mathbf{S}(\bar{\mathbf{D}}_{6}+\rho\tilde{\mathbf{D}}_{6})\mathbf{S}^{H} + \mathbf{D}_3^{H}(\bar{\mathbf{D}}_8+\rho\tilde{\mathbf{D}}_8)\mathbf{D}_3+(\rho/\sigma^2_d)\mathbf{D}_3^H\mathbf{D}_3 \Big\}\boldsymbol{\Psi}_{\text{T}}.
\end{align}}
For a fixed channel realization, the SINR of the $k$th symbol can now be obtained by substituting $\mathbf{R}_v$ in \eqref{eq:PP-SINR_ICSI_fianl}. We see that the SINR expression in \eqref{eq:PP-SINR_ICSI_fianl} calculates  $\mathbf{S}=\mathbf{D}^{-1}_{\text{LM}}=(\mathbf{D}^{H}\mathbf{D}+\rho\mathbf{I}_{N_tMN})^{-1}$. This inverse can be calculated using the proposed Algorithm-\ref{algo:DA_INV} with a computational complexity of $\mathcal{O}(MN)$,  as derived in \emph{Lemma} \ref{Lemma2_1}. Multiplication with matrices $\boldsymbol{\Psi}_{\text{T}}$ and $\boldsymbol{\Psi}_{\text{T}}^{H}$ can be performed,  as explained shortly in the paragraph below \eqref{eq:y_tilde}, with $\mathcal{O}(MN\mbox{log}_2MN)$ complexity. SINR calculation in \eqref{eq:PP-SINR_ICSI_fianl} for the LM receiver has thus $\mathcal{O}(MN)+\mathcal{O}(MN\mbox{log}_2MN)$ complexity.    

\subsection{SINR calculation for the proposed LZ receiver with imperfect receive CSI}\label{MMSE_ZF}
The LZ receiver with  CSI error, by substituting \eqref{eq:Decom_Dhat} in \eqref{eq:Combiners1},  is given as follows
\begin{equation}
\widehat{\mathbf{G}}_{\text{LZ}}^{H} = \boldsymbol{\Psi}_{\text{T}}^H(\mathbf{D}+\Delta\mathbf{D})^{\dagger}\boldsymbol{\Psi}_{\text{R}}.
\end{equation}
Here $(\cdot)^{\dagger}$ is the pseudo-inverse. By expanding $(\mathbf{D}+\Delta\mathbf{D})^{\dagger}$ using first-order Taylor series \cite{petersen2012matrix}:
\begin{equation}\label{eq:ZF_approx}
\widehat{\mathbf{G}}^H_{\text{LZ}} \cong \boldsymbol{\Psi}_{\text{T}}^H\mathbf{D}^{\dagger}(\mathbf{I}_{N_rMN}-\Delta\mathbf{D}\mathbf{D}^{\dagger})\boldsymbol{\Psi}_{\text{R}}.
\end{equation}
The LZ estimate of transmit vector $\mathbf{x}$, with CSI errors, is now calculated using \eqref{eq:mimo_otfs} and \eqref{eq:ZF_approx} as
\begin{align}\label{eq:ZF_Est}
\hat{\mathbf{x}}_{\text{LZ}} &= \boldsymbol{\Psi}_{\text{T}}^H\mathbf{D}^{\dagger}(\mathbf{I}_{N_rMN}-\Delta\mathbf{D}\mathbf{D}^{\dagger})\boldsymbol{\Psi}_{\text{R}}(\mathbf{H}\mathbf{x}+\tilde{\mathbf{v}})\stackrel{(a)}{=}\mathbf{x}+\bar{\mathbf{v}}.
\end{align}
Equality in (a) follows by substituting the decomposition of $\mathbf{H}$ from \eqref{eq:mimo_otfs_mtx_D}. {The vector $$\bar{\mathbf{v}}=-\boldsymbol{\Psi}_{\text{T}}^H\mathbf{D}^{\dagger}\Delta\mathbf{D}\boldsymbol{\Psi}_{\text{T}}\mathbf{x}+\boldsymbol{\Psi}_{\text{T}}^H\mathbf{D}^{\dagger}\boldsymbol{\Psi}_{\text{R}}\tilde{\mathbf{v}}-\boldsymbol{\Psi}_{\text{T}}^H\mathbf{D}^{\dagger}\Delta\mathbf{D}\mathbf{D}^{\dagger}\boldsymbol{\Psi}_{\text{R}}\tilde{\mathbf{v}}$$ is the noise-plus-distortion caused by channel estimation error matrix $\Delta\mathbf{H}$.} The SINR of the $k$th symbol, using \eqref{eq:ZF_Est}}, is given as
\begin{align}\label{eq:SINR_ZF}
\gamma^k_{\text{LZ}} = \dfrac{P_x}{[\mathbf{R}_{\bar{v}}]_{(k,k)}}, \ \ \text{for}\ \ k=1,2,\ldots,N_tMN.
\end{align}
Using the property that the noise vector $\mathbf{v}$ and the error matrix $\Delta\mathbf{D}$ are  independent with zero mean\cite{WangAMMCL07}, the covariance matrix $\mathbf{R}_{\bar{v}}=\mathbb{E}[\bar{\mathbf{v}}\bar{\mathbf{v}}^H]$, for a fixed channel realization, is given as
\begin{align}\label{eq:Cov_w1}
\nonumber \mathbf{R}_{\bar{v}}&= \boldsymbol{\Psi}_{\text{T}}^H\Big\{\sigma^2_v\mathbf{D}^{-1}_{\text{LZ}}+P_x\mathbf{D}^{-1}_{\text{LZ}}\mathbf{D}^H\mathbb{E}\big[\Delta\mathbf{D}\Delta\mathbf{D}^H\big]\mathbf{D}\mathbf{D}^{-1}_{\text{LZ}}\\
&+\sigma^2_v\mathbf{D}^{-1}_{\text{LZ}}\mathbf{D}^H\mathbb{E}\big[\Delta\mathbf{D}\mathbf{D}^{-1}_{\text{LZ}}\Delta\mathbf{D}^H\big]\mathbf{D}\mathbf{D}^{-1}_{\text{LZ}}\Big\}\boldsymbol{\Psi}_{\text{T}}.
\end{align}
Recall that $\mathbf{D}_{\text{LZ}}=\mathbf{D}^H\mathbf{D}$. Using \emph{Lemma} \ref{Lemma3}, we get $\mathbb{E}\big[\Delta\mathbf{D}\Delta\mathbf{D}^H\big]=N_t\sigma^2_d\mathbf{I}_{N_rMN}$. Exploiting this result and \emph{Lemma} \ref{Lemma4}, the final expression for $\mathbf{R}_{\bar{v}}$ can be obtained from \eqref{eq:Cov_w1} as
\begin{align}\label{eq:Cov_w_final}
\mathbf{R}_{\bar{v}}= P_x\sigma^2_d\boldsymbol{\Psi}_{\text{T}}^H &\bigg\{\rho\mathbf{D}^{-1}_{\text{LZ}}\mathbf{D}^H\bar{\mathbf{D}}_{\text{LZ}}\mathbf{D}\mathbf{D}^{-1}_{\text{LZ}}+\dfrac{\left(\rho+N_t\sigma^2_d\right)}{\sigma^2_d}\mathbf{D}^{-1}_{\text{LZ}}\bigg\}\boldsymbol{\Psi}_{\text{T}}.
\end{align}
Here the matrix $\bar{\mathbf{D}}_{\text{LZ}}\in\mathbb{C}^{N_rMN\times N_rMN}$ is $\bar{\mathbf{D}}_{\text{LZ}} = \sum_{i=1}^{N_t}\big(\mathbf{I}_{N_r}\otimes \left[\mathbf{D}^{-1}_{\text{LZ}}\right]_{i,i}\big)$. Finally, the SINR for the $k$th symbol of the  LZ receiver is obtained by substituting $[\mathbf{R}_{\bar{v}}]_{(k,k)}$ in \eqref{eq:SINR_ZF}. Note that the inverse $\mathbf{D}^{-1}_{\text{LZ}}$ in \eqref{eq:Cov_w_final} can be obtained using Algorithm-\ref{algo:DA_INV} with a computational complexity of $\mathcal{O}(MN)$. Furthermore,  multiplication with the matrices $\boldsymbol{\Psi}_{\text{T}}$ and $\boldsymbol{\Psi}_{\text{T}}^{H}$ has $\mathcal{O}(MN\mbox{log}_2MN)$ complexity. Thus, similar to the SINR of the LM receiver, complexity for calculating the SINR of the proposed LZ receiver in \eqref{eq:PP-SINR_ICSI_fianl} is also $\mathcal{O}(MN)+\mathcal{O}(MN\mbox{log}_2MN)$.

With the SINR $\gamma^{k}_{\text{A}}$, for $\text{A}\in\{\text{LZ, LM}\}$, in \eqref{eq:PP-SINR_ICSI_fianl} and  \eqref{eq:SINR_ZF} for the LM and LZ receivers respectively, their BER with imperfect CSI can now be found by plugging $\gamma^{k}_{\text{A}}$ in the following expression $k_{0}Q\Big(\sqrt{k_{1}\gamma^{k}_{\text{A}}}\Big)$, where constants $k_{0}$ and $k_{1}$ depend on the constellation order. 

%For example, the BER for the MIMO-OTFS system relying on $4$-QAM constellation can be computed as~\cite{lin2010filter}
%\begin{eqnarray}\label{eq:BER_QAM_ICSI_4}
%{ \big[\text{BER}|\gamma_{\text{A}}\big]_{4}^{k} \simeq 2Q\left(\sqrt{\gamma^{k}_{\text{A}}}\right).}
%\end{eqnarray}
%We can similarly compute BER for higher-order QAM constellations by substituting $\gamma^{k}_{\text{A}}$ and appropriate $k_{0}$ and $k_{1}$ values \cite{lin2010filter}.

\section{Simulation results}\label{Results}
We now investigate the BER of  proposed designs for a spatially-multiplexed  MIMO-OTFS system with an ideal pulse which satisfies the bi-orthogonality property. Such ideal pulses are commonly used in OTFS systems \cite{SurabhiC20,SurabhiAC19}. We consider two  MIMO configurations viz  $N_r\times N_t=4\times 2$ and $N_r\times N_t=4\times 4$, with either BPSK or QPSK constellation. We assume the number of delay bins $M$ and the Doppler bins $N$  as $M=N=32$, carrier frequency of $4$ GHz, and a subcarrier spacing of $15$ KHz. We use a $5$-tap delay-Doppler channel with its parameters given in Table-~\ref{table:2}, and define SNR as $P_x/\sigma^2_v$ . 
%  \begin{table}[h!]
%\centering
%\caption{Simulation Parameters}
%\begin{tabular}{| m{12em} | m{3cm}|} 
% \hline
% Parameter & Value  \\ 
% \hline
% No. of subcarriers (M) & 32 \\ 
% \hline
% No. of OTFS symbols (N) & 32  \\
% \hline 
%Carrier frequency & 4 GHz  \\
%\hline 
% Subcarrier spacing & 15 KHz  \\
%  \hline
%  Constellation  & BPSK and QPSK  \\
%  \hline
%  MIMO Size  & $2\times 2$ and $4\times 4$ \\
%  \hline
%\end{tabular}
%\label{table:1}
%\end{table}
\begin{table}[h!]
\centering
\caption{\footnotesize Delay-Doppler channel parameters}
\begin{tabular}{| m{11em} | m{.5cm}| m{1cm}|  m{1cm}| m{1cm}| m{1cm}|} 
 \hline
 Channel tap no. & 1 & 2 & 3 & 4 & 5   \\ 
 \hline
 Delay ($\mu s$) & 2.08 & 5.20 & 8.328 & 11.46 & 14.80 \\ 
 \hline
 Doppler shift (Hz) & 0 & 470 & 940 & 1410 & 1851  \\
 \hline 
Power of channel tap (dB) & 1 & -1.804 & -3.565 & -5.376 & -8.860  \\
  \hline
\end{tabular}
\label{table:2}
\end{table}
For each SNR point,  BER is numerically  computed by averaging over $100$ i.i.d. channel realizations. We abbreviate the i) conventional ZF/MMSE  as cZF/cMMSE; ii) perfect and imperfect receive CSI scenarios as PCSI and ICSI, respectively.

\textbf{BER comparison of  conventional and proposed ZF and MMSE receivers:}
Fig.~\ref{fig:BER1}(a) and Fig.~\ref{fig:BER1}(b) show the BER of conventional and proposed ZF/MMSE receivers for a MIMO-OTFS system with QPSK constellation, integer Doppler shift, and with perfect and imperfect receive CSI, respectively. For imperfect receive CSI, channel estimation error variance $\sigma_e^2$ in \eqref{eq:CFR_EST} is set as $\rho/N_t$ \cite{HassibiH03}. We see that for both MIMO configurations, BER of the proposed LZ and LM receivers exactly
match their conventional counterparts. This is because the proposed designs do not make any approximation and exploit the following inherent properties i) doubly-circulant structure of the MIMO-OTFS channel matrix $\mathbf{H}$ by decomposing it  as shown in Eq. (11) as $\mathbf{H}=  \boldsymbol{\Psi}_{\text{R}}^H\mathbf{D}\boldsymbol{\Psi}_{\text{T}}$; ii) 
fact that $\mathbf{D}\in \mathcal{C}_{N_rN_t,MN}$ (set of block matrices with $N_rN_t$ blocks of $MN\times MN$ diagonal matrices); and iii)  block-wise inverse of matrices and \emph{Schur} Complement.   \textit{{We note that it is not unusual for a design to have lower complexity without degrading in performance \cite{SurabhiC20}.}}  

 \begin{figure}[htbp]
	\begin{center}
		\subfloat[]{\includegraphics[scale = 0.52]{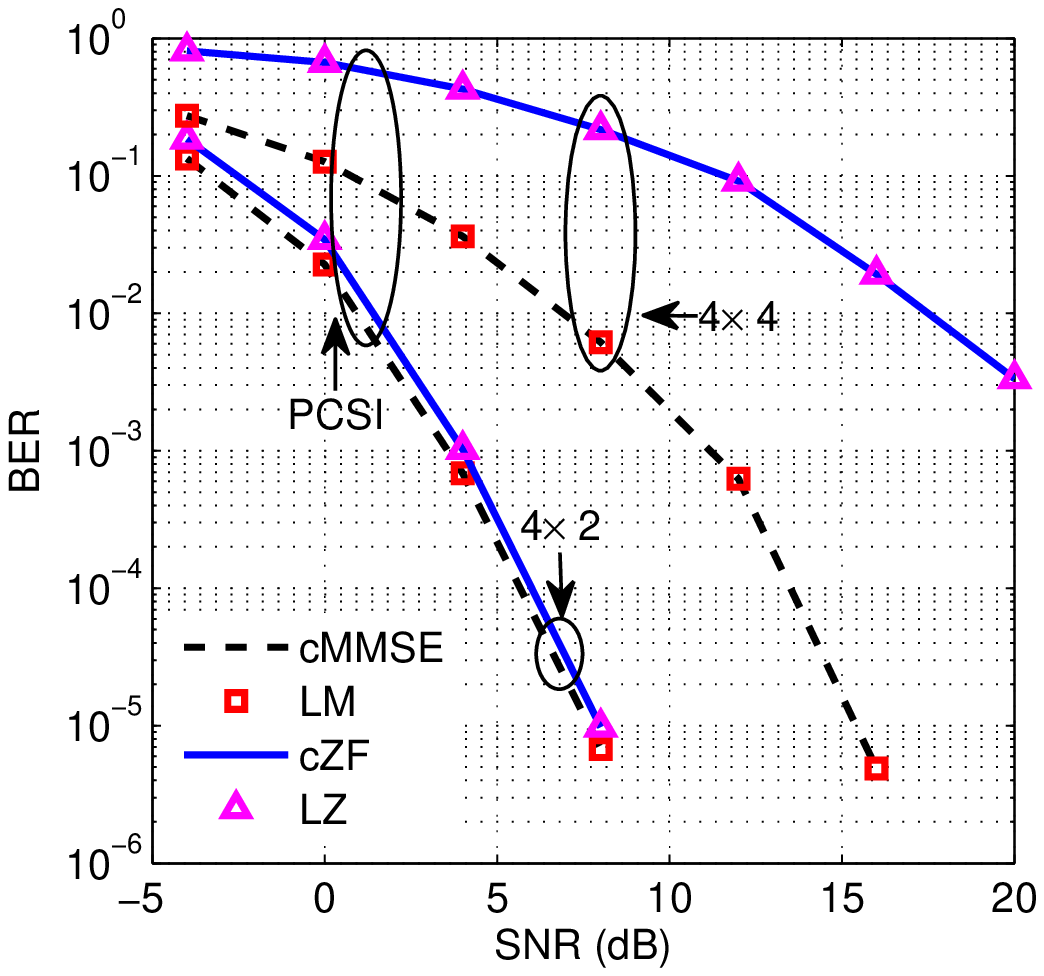}}
		\hfil 
		\subfloat[]{\includegraphics[scale = 0.52]{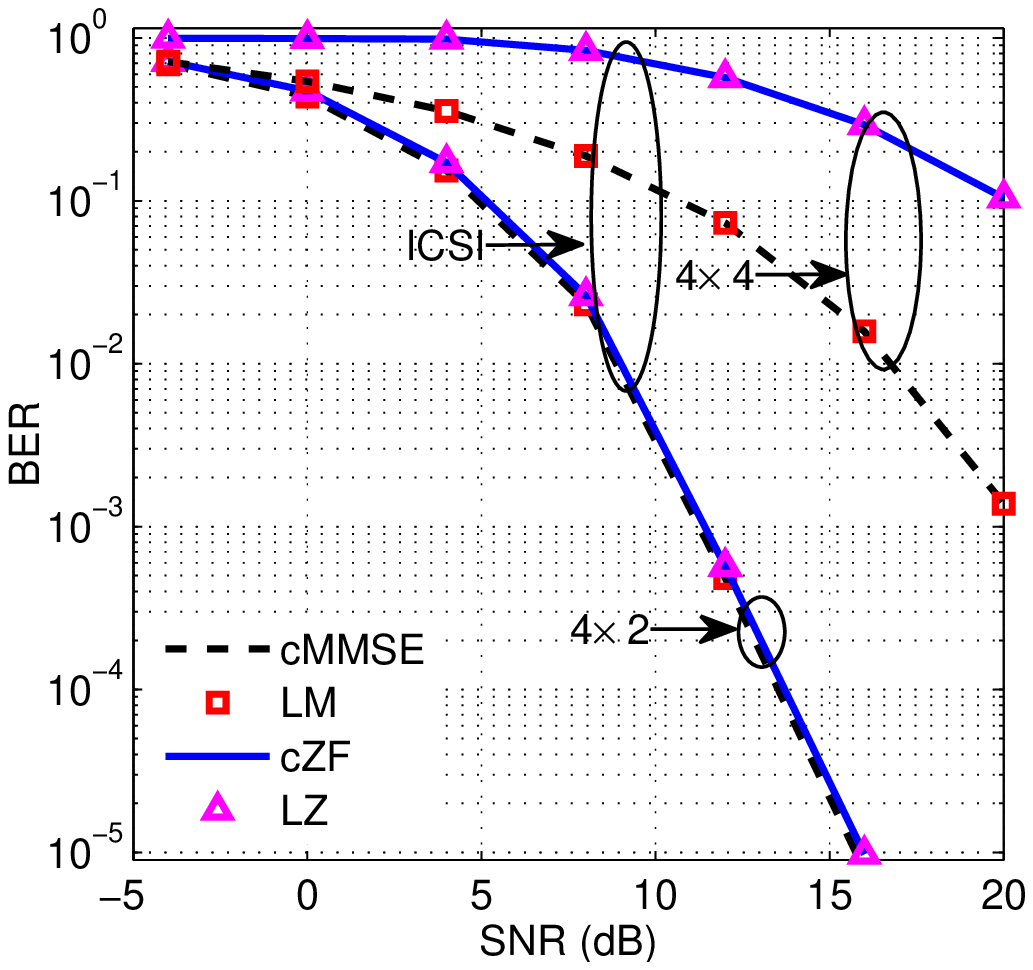}}
		\hfil 
		\subfloat[]{\includegraphics[scale = 0.53]{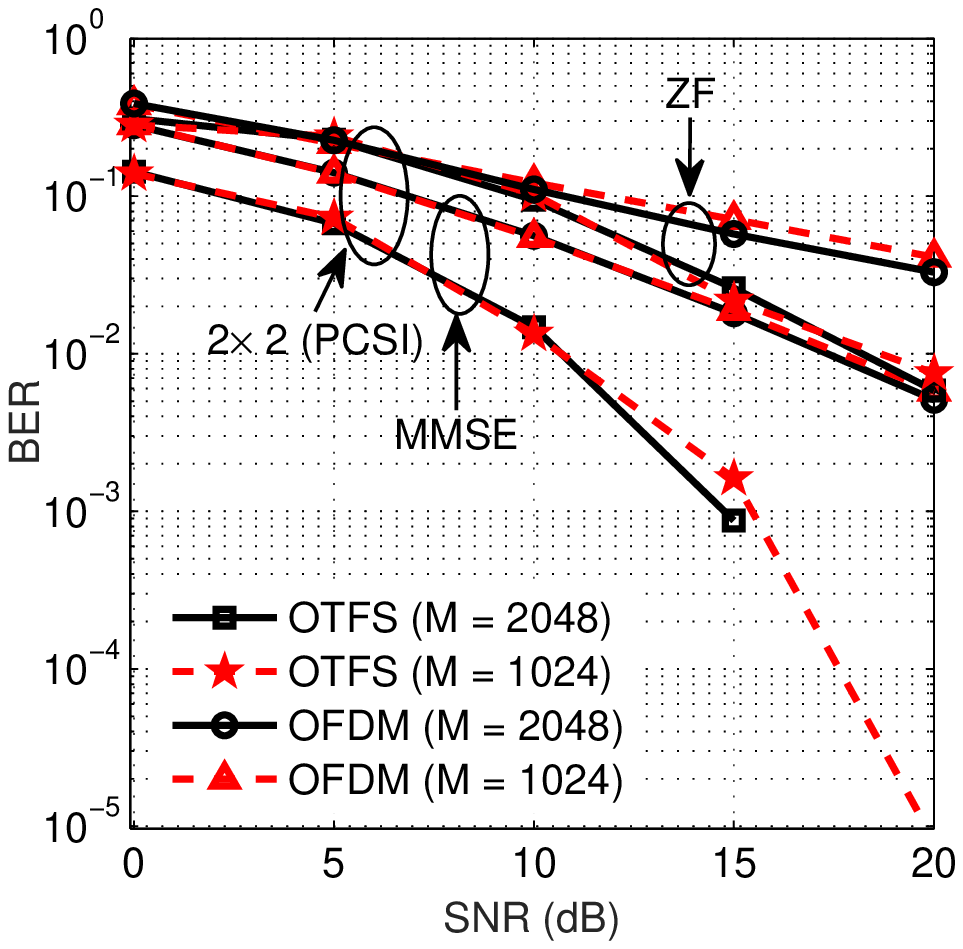}}
		
	\end{center}
	\hfil
	\caption{\footnotesize BER of the proposed LZ and LM receivers and their conventional counterparts  for $4\times 2$ and $4\times 4$ MIMO-OTFS systems with integer Doppler and QPSK modulation for (a) perfect CSI; and (b) Imperfect CSI. {(c) BER  of OTFS (integer Doppler) and MIMO-OFDM systems with $N_t=N_r=2$, $N=6$, $M\in\{1024,2048\}$, QPSK modulation and perfect CSI.}}
	\label{fig:BER1}
\end{figure}

Fig.~\ref{fig:BER1}(c) compares the BER of OTFS- and OFDM-based MIMO systems with $M=1024$ and $M=2048$. The model in \cite{RottenbergMHL19} is used to implement the OFDM-MIMO system for time-varying channels.  We see that the MIMO-OTFS systems have significantly lower BER. This is because the former, due to delay-Doppler domain transmission, is robust to time-varying channels. We again observe that the proposed receivers BER match their conventional counterparts.

\begin{figure}[htbp]
	\begin{center}
		\subfloat[]{\includegraphics[scale = 0.75]{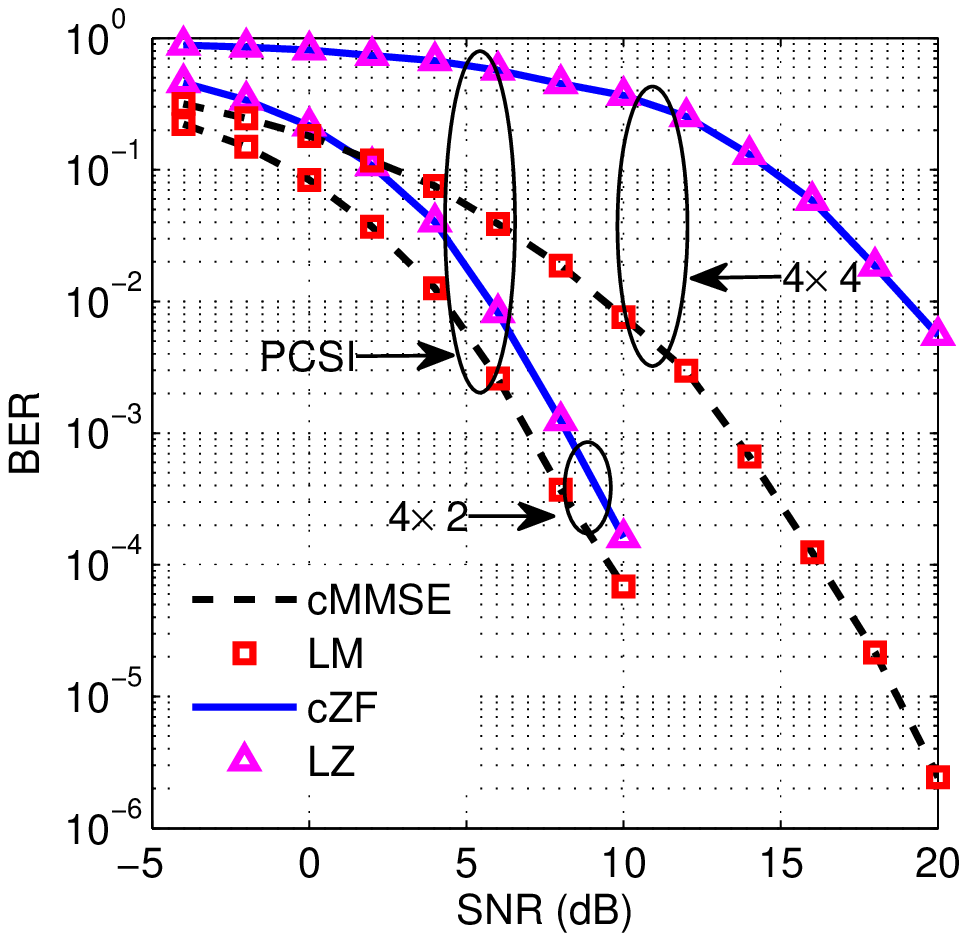}}
		\hfil 
		\hspace{-10pt}\subfloat[]{\includegraphics[scale = 0.75]{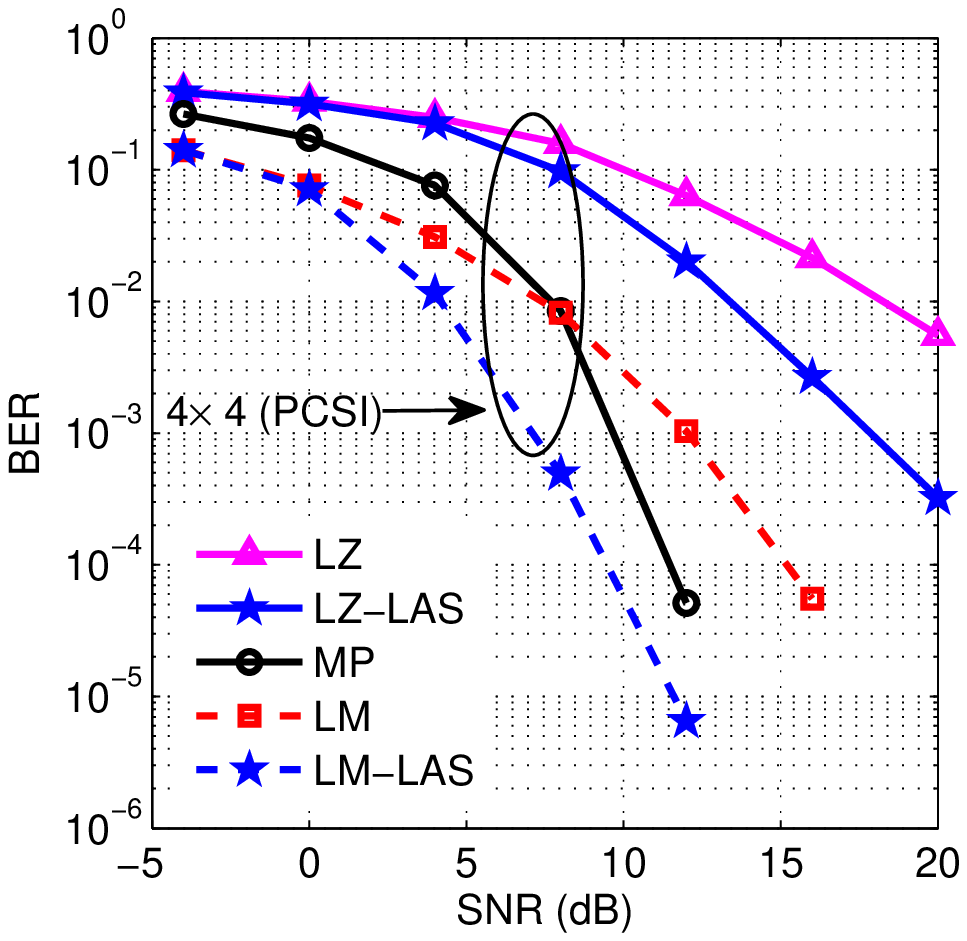}}
		\hfil
	\end{center}
	\caption{\footnotesize (a) BER comparison of the proposed LZ and LM with the conventional receivers for $4\times 2$ and $4\times 4$ MIMO-OTFS systems with QPSK and fractional Doppler. (b) BER comparison of proposed LZ, LM receivers, with and without LAS technique,  and MP receiver for $4\times 4$ MIMO-OTFS systems with BPSK  and integer Doppler.}
	\label{fig:BER2}
\end{figure}
 
%\vspace{-0.1cm}

%These numerical studies clearly show that the BER of the proposed low-complexity ZF and MMSE receivers exactly predict the corresponding BER of the conventional ZF and MMSE receivers.

 Fig.~\ref{fig:BER2}(a) compares the BER of the proposed  LZ and  LM receivers and their conventional counterparts for a fractional Doppler shift instead of the integer one,   the rest of system settings being same. We observe that the behavior is similar to that of Fig.~\ref{fig:BER1}(b). Fig.~\ref{fig:BER2}(b) compares the BER of proposed receivers and the widely-used non-linear MP-based receiver for a $4\times 4$ MIMO-OTFS system with BPSK modulation, and perfect receive CSI. For a fair comparison with non-linear MP receiver, we also plot the BER of LZ-LAS and LM-LAS receivers wherein the proposed LZ and LM receivers are followed by low-complexity  local-search-based non-linear likelihood ascent search (LAS) operation. The LAS receiver begins with an initial solution provided by the LZ or LM receiver, and searches for good solutions in the neighborhood until a local optimum is reached \cite{chockalingam2014large}. We see that the   i) proposed LM receiver comprehensively outperforms the MP receiver at the low SNR; ii) LM-LAS receiver has  significantly lower BER than the MP receiver for all SNR values.  The complexity of the LM-LAS scheme, as shown in the next subsection,  is almost similar to that of the proposed LM  design.
%\colr{1) use subfigure instead of subfloat - gives better control. Ask Ekant.} \colb{Sir, I tried using subfigure. It is giving some error that I am not able to figure out yet. I will fix it soon.}

%\begin{figure*}[t!]
%\centering
%}{\begin{subfigure}[b]{.49\linewidth}}{\begin{subfigure}[b]{0.32\linewidth}}
%   \includegraphics[width=\linewidth]{SC_FBMC_IMCSI_AR_Vs_nRx_10dB.eps}
%   \caption{\small }
%\label{fig:SC_AR_Vs_nRx_10dB}
%\end{subfigure}
%{\begin{subfigure}[b]{.49\linewidth}}{\begin{subfigure}[b]{0.32\linewidth}}
%   \includegraphics[width=\linewidth]{BER_CON_Vr_LC_ZF_MMSE_QPSK_4C2_4C4_PCSI_FD_COMP.eps}
%   \caption{\small }
%\label{fig:SC_AR_Vs_Pu}
%\end{subfigure}
%\caption{Uplink sum-rate versus  a) number of BS antennas for perfect and imperfect CSI with the transmit power per user $2P_{d}=10$ dB; and b) transmit power per user for different number of BS antennas with imperfect CSI.}\label{fig:SC_COMP} 
%\end{figure*}

\textbf{Complexity comparisons:}
We now compare the computational complexity of the proposed LZ and LM receivers with  the conventional MMSE and MP receivers \cite{RavitejaPH19,RamachandranC18}. Total number of operations for the proposed  LZ and LM receivers, as shown in \eqref{eq:overall_complexity_MMSE},  are related as $\mu_{\text{LM}}=\mu_{\text{LZ}}+2N_t MN$.  The proposed designs, therefore, have similar complexities. Conventional ZF and MMSE receiver in MIMO-OTFS, as mentioned in Section-\ref{complexity1},  have  $\mathcal{O}(N_t^3M^3N^3)$ complexity. As the LAS operation has $\mathcal{O}(N_tMN)$ complexity, the LM-LAS receiver has $\mu_{\text{LM}}+\mathcal{O}(N_tMN)$ complexity. The MP receiver complexity in MIMO-OTFS systems, as discussed in Section-\ref{complexity},  varies as $\mathcal{O}(N_IN_rN_tMNSQ)$, where $S=\sum_{i=1}^{L_h}(2N_i+1)$ with $N_i=10$ \cite{raviteja2018interference}.

 \begin{figure}[htbp]
 \begin{center}
	\subfloat[]{\includegraphics[scale = 0.54]{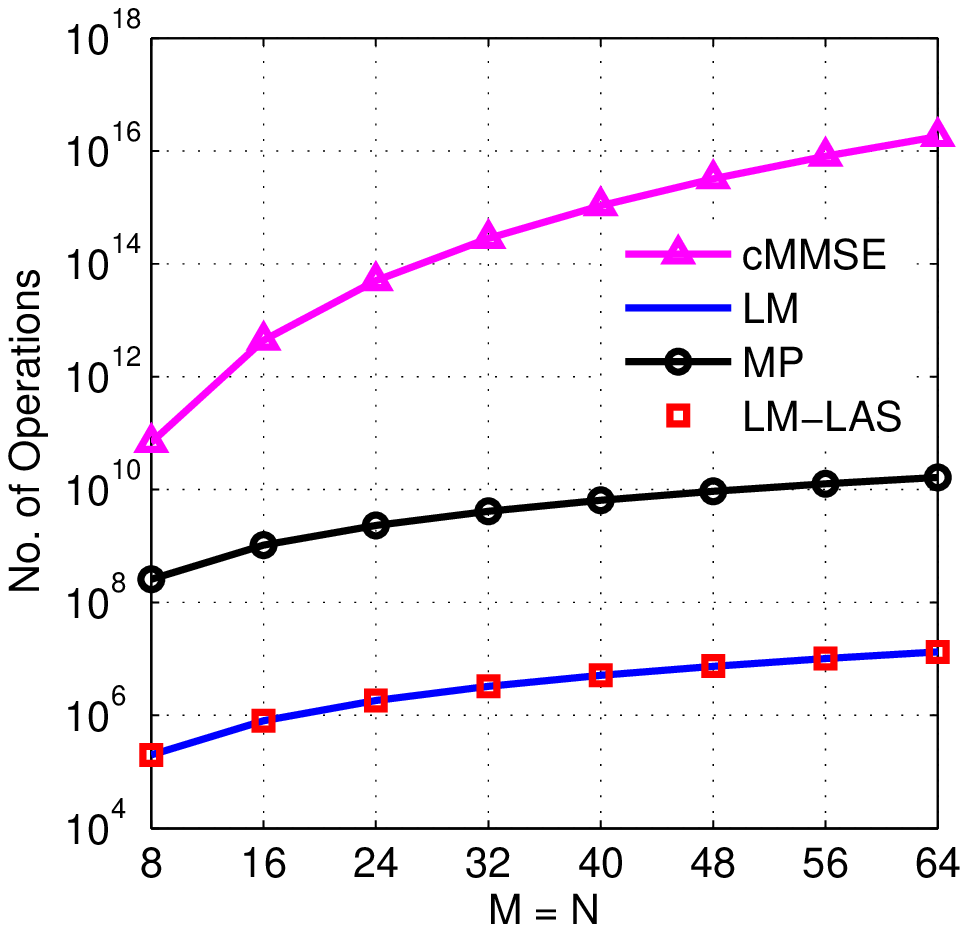}}
	\hfil 
	\hspace{-10pt}\subfloat[]{\includegraphics[scale = 0.54]{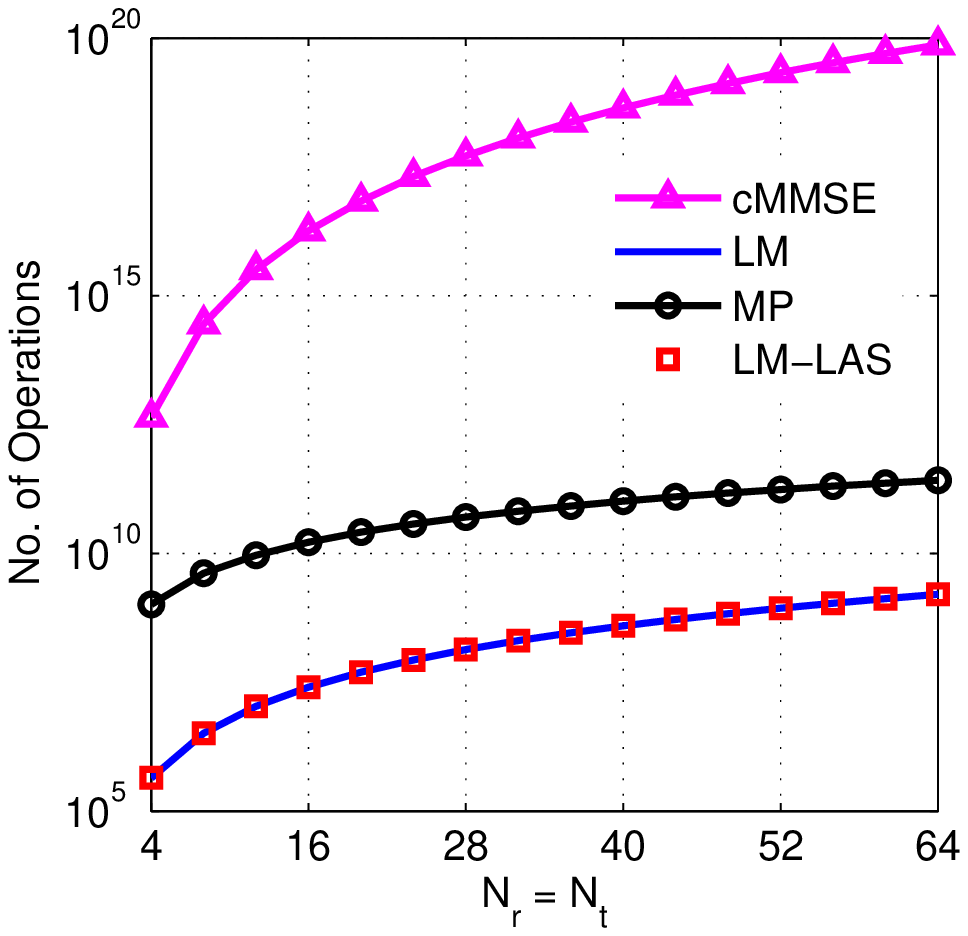}}
	\hfil
	\hspace{-10pt}\subfloat[]{\includegraphics[scale = 0.54]{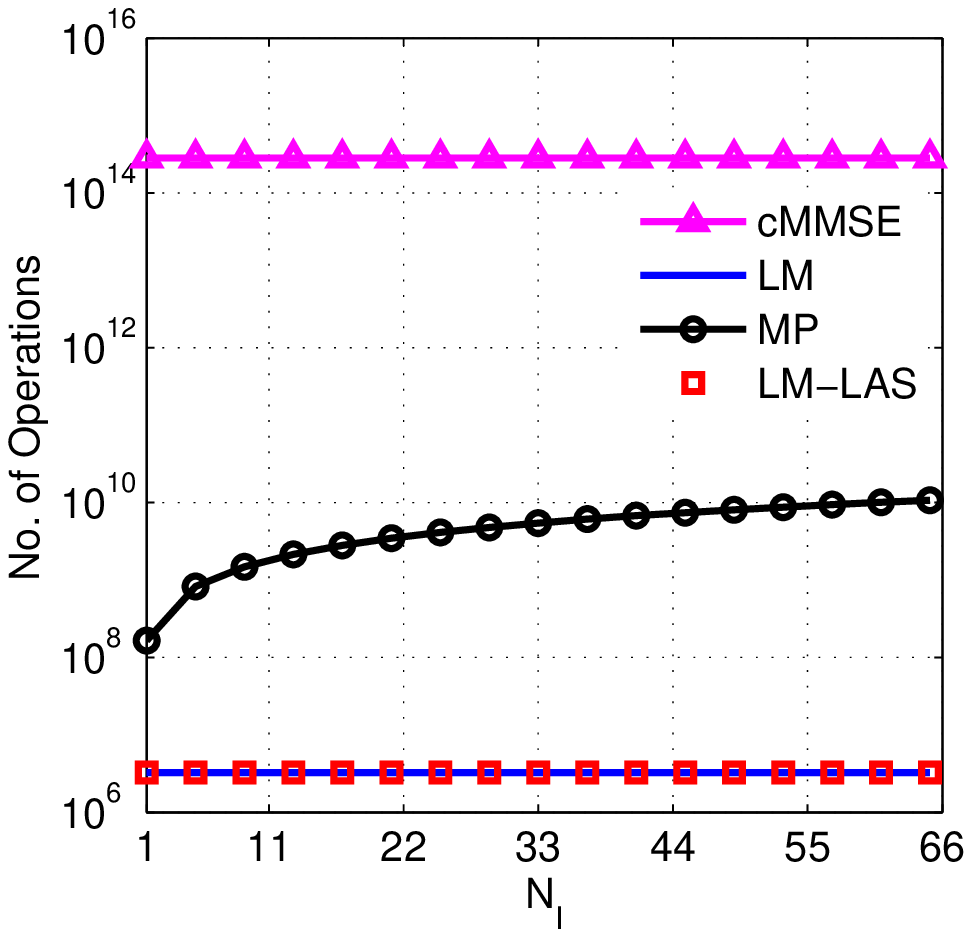}}
\end{center}
	\hfil
	\caption{\footnotesize Complexity comparison of the proposed  and the existing techniques: (a) number of operation versus $M=N$ with $N_r=N_t=8$ and $N_I=20$ iterations for the MP receiver; (b) number of operation versus the number of receive antennas $N_r=$ the number of transmit antennas $N_t$ with $M=N=32$ and $N_I=20$ iterations for the MP receiver; and (c) number of operation versus the number of iterations $N_I$ with $M=N=32$ and $N_r=N_t=8$. }%\colr{shall we plot the order or talk in terms of flops} \colb{Sir, We can also plot flops by simply multiplying the number operations by a constant that converts number of operation to flops. Since paper by Prof. Chockalingam also plotted number of operation, we followed the same pattern.}
	\label{fig:complexity}
\end{figure}

Fig.~\ref{fig:complexity}(a) shows the number of operations for varying $M=N$, where $M$ and $N$ are the number of delay and Doppler bins, respectively. We use $N_I=20$ iterations to evaluate MP receiver complexity, within which it typically converges \cite{raviteja2018interference}. We see that the  complexity of LM and LM-LAS receivers is almost similar. Their complexity is, however, significantly lower than the MP and  conventional MMSE receivers. This is because the proposed designs exploit doubly-circulant channel structure.  \textit{We also note that it is not unusual for a design to have lower complexity without degrading the performance \cite{SurabhiC20}.}
We see a similar behavior in Fig.~\ref{fig:complexity}(b), where we plot the number of operations versus $N_r=N_t$ antennas.% \colr{explain relation between complexity order and number of operations} \colb{Sir, both are number of operations. The only difference is that we know exact number of operation for the proposed techniques, however, for the conventional ML and MP schemes, we know only the order of operations from the literature.}.

Fig.~\ref{fig:complexity}(c) shows the complexities of four receivers as a function of number of iteration $N_I$. For this study, we fix $M=N=32$  and $N_r=N_t=8$ antennas. We once again observe that the complexity of the proposed LM and LM-LAS receivers is significantly lower than the conventional MMSE and the MP receivers. Since the proposed and conventional MMSE receivers are not iterative, their complexities remain constant with $N_I$. Understandably, the computational cost of the MP receiver, due to its iterative nature, increases with $N_I$.

\begin{figure}[htbp]
	\begin{center}
	\subfloat[]{\includegraphics[scale = 0.75]{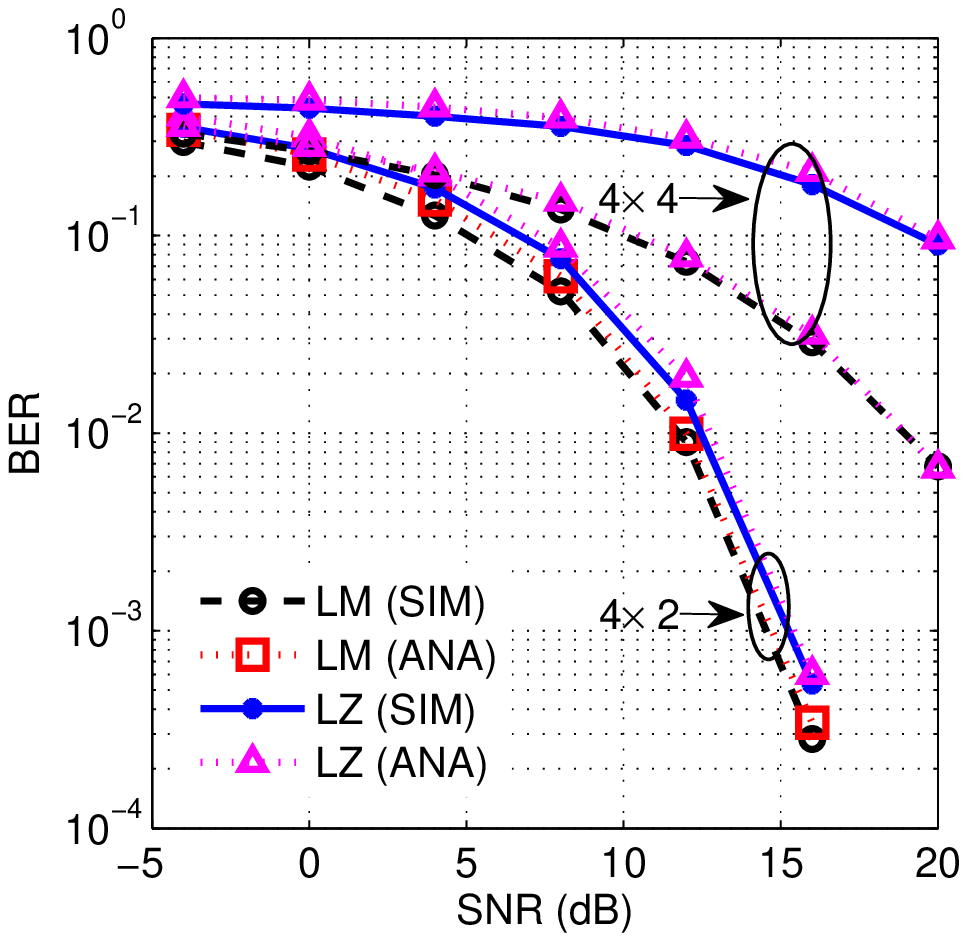}}
	\hfil 
	\hspace{-10pt}\subfloat[]{\includegraphics[scale = 0.75]{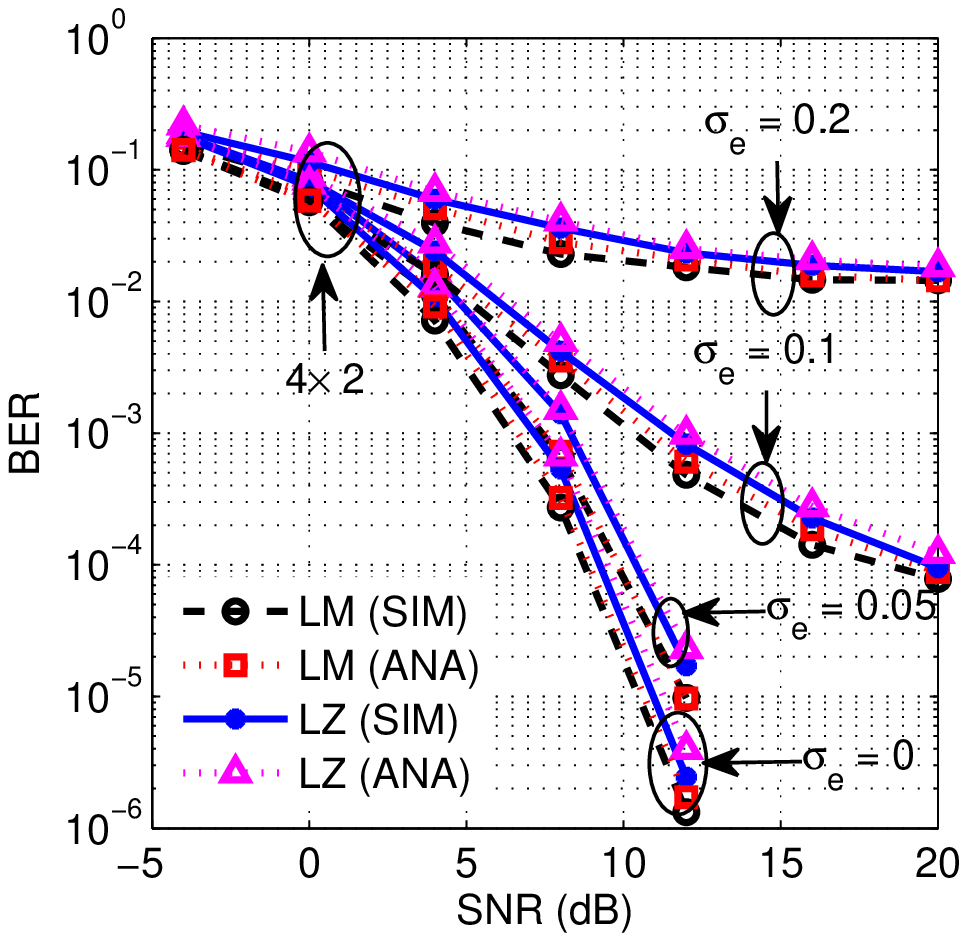}}
	\end{center}	
	\hfil
	\caption{\footnotesize BER versus SNR comparison of the simulated and analytical expressions for the proposed LZ and LM receivers with integer Doppler and QPSK modulation: (a) for $4\times 2$ and $4\times 4$ MIMO-OTFS systems with $\sigma^2_e=\rho/N_t$; and (b) $4\times 2$ MIMO-OTFS systems with $\sigma^2_e$ being independent for the SNR.}
	\label{fig:BER3}
\end{figure}

\textbf{Analytical and simulated BER comparisons:}
We now validate the BER expressions derived in  Section-\ref{BER} for the proposed LZ and LM receivers with imperfect CSI. For this study, we first  make the channel estimation error  variance  $\sigma^2_e$  a function of the SNR and vary it  as $\sigma^2_e=\rho/(N_t)$ \cite{HassibiH03}, where $\rho=\sigma^2_v/P_x$ = 1/SNR. We later make $\sigma^2_e$, and  hence $\sigma^2_d=\sigma^2_e\sum_{i=1}^{M}L_h^i$, independent of SNR. 
Fig.~\ref{fig:BER3}(a) shows analytical and simulated BER for the proposed receivers with channel estimation error variance $\sigma^2_e=\rho/(N_t)$. We see that the analytical BER of both the proposed receivers, derived using corresponding SINR expressions in \eqref{eq:PP-SINR_ICSI_fianl} and \eqref{eq:SINR_ZF}, closely match the simulated ones. This study validates that the derived SINR expressions accurately model the BER of ZF and MMSE receivers for MIMO-OTFS systems. Fig.~\ref{fig:BER3}(b) compares analytical and simulated BER values for $4\times 2$ MIMO-OTFS system with the same system settings as in Fig.~\ref{fig:BER3}(a) but $\sigma^2_e$ now being independent of SNR. The analytical and simulated counterparts again match. Also, the channel estimation error dominates in  high SNR, which is not surprising.%, since it is hidden by the noise in the low SNR range and shows up in the weak noise regime (i.e. high SNR).

%In the plots, $\sigma^2_e=0$ corresponds to the perfect receive CSI scenario, because, with $\sigma^2_e=0$ (also implies $\sigma^2_d=0$) \colr{it is a good practice to mention what $\sigma^e_d$ and $\sigma^2_d$ at multiple places}, the SINR expressions for the proposed ZF and MMSE receivers in \eqref{eq:SINR_ZF} and \eqref{eq:PP-SINR_ICSI_fianl} respectively reduce to the scenario perfect receive CSI.

\section{Conclusions}\label{Conclusion}
%\vspace{-0.2cm}
We proposed  novel low-complexity ZF (LZ) and MMSE (LM) receivers, which  exploit inherent characteristics of MIMO-OTFS channel and the properties of block matrices, to achieve lower complexity than the conventional ZF, MMSE and  message passing (MP) receivers.  We showed that the BER of i)  both LZ and LM receivers, for perfect and imperfect receive CSI, have exactly the same BER as that of their conventional counterparts; and  ii) LM receiver at low SNR values is lower than the non-linear MP receiver. The LM receiver, when combined with non-linear likelihood ascent search technique, outperforms the MP receiver at all SNR values. We derived analytical BER expressions for  LZ and LM receivers and showed that, when averaged over multiple channel realizations, they precisely match their respective simulated BERs. {The current receivers did not exploit OTFS channel sparsity. Future work can exploit  circulant channel structure and its sparsity to further reduce the complexity.} 
{Future work can also develop low complexity receivers by considering practical pulses. The  low-complexity receivers proposed herein with ideal pulse can serve as a starting point for design of their counterparts with non-ideal pulse, and benchmark their performance.}
%{Future work can also develop low complexity algorithms by considering practical pulses. The  low-complexity receivers proposed herein with ideal pulse can benchmark their performance.}
\appendices

%\section{Proof of Lemma-\ref{Lemma2}}
\section{{}}
\label{Lemma2_proof}
To begin with, let $t=N_t$. The matrix $\mathbf{X}$, similar to \eqref{eq:D_A}, can always be partitioned into four sub-matrices, namely $\mathbf{A}_{N_t-1}\in\mathbb{C}^{(N_t-1)MN\times (N_t-1)MN}$, $\mathbf{B}_{N_t-1}\in\mathbb{C}^{(N_t-1)MN\times MN}$, $\mathbf{C}_{N_t-1}\in\mathbb{C}^{MN\times (N_t-1)MN}$ and $\mathbf{D}_{N_t-1}\in\mathbb{C}^{MN\times MN}$. With the above partitioning of $\mathbf{X}$, it follows from \eqref{eq:mimo_otfs_mtx1},  that $\mathbf{X}^{-1}$ can be computed block-wise, provided the matrix $\mathbf{D}_{N_t-1}$ and its \emph{Schur} complement $\mathbf{S}_{N_t-1}=\mathbf{A}_{N_t-1}-\mathbf{B}_{N_t-1}\mathbf{D}^{-1}_{N_t-1}\mathbf{C}_{N_t-1}$ are invertible. Since $\mathbf{X}\in\mathcal{C}_{N_t^2,MN}$, the sub-matrix $\mathbf{D}_{N_t-1}$ is always an $MN\times MN$ diagonal matrix {with all its elements are $> 0 $ \cite{SurabhiC20}}. Thus, $\mathbf{D}^{-1}_{N_t-1}$ always exists. For $\mathbf{X}^{-1}$ to exist, we have to next prove that $\mathbf{S}_{N_t-1}^{-1}$ exists. We see that from \emph{Lemma} \ref{Lemma1}, the \emph{Schur} complement $\mathbf{S}_{N_t-1}\in\mathcal{C}_{(N_t-1)^2,MN}$. This is because all the matrices  $\mathbf{A}_{N_t-1}$, $\mathbf{B}_{N_t-1}$, $\mathbf{C}_{N_t-1}$ and $\mathbf{D}_{N_t-1}$ comprise blocks of $MN\times MN$ diagonal matrices.  Thus, similar to \eqref{eq:D_A}, $\mathbf{S}^{-1}_{N_t-1}$ can be calculated block-wise by partitioning $\mathbf{S}_{N_t-1}$ in terms of the sub-matrices $\mathbf{A}_{N_t-2}$, $\mathbf{B}_{N_t-2}$, $\mathbf{C}_{N_t-2}$ and $\mathbf{D}_{N_t-2}$, and by imposing the conditions that $\mathbf{D}_{N_t-2}$ and its \emph{Schur} complement $\mathbf{S}_{N_t-2}=\mathbf{A}_{N_t-2}-\mathbf{B}_{N_t-2}\mathbf{D}^{-1}_{N_t-2}\mathbf{C}_{N_t-2}$ are invertible. Proceeding in this way, let $\mathbf{S}_{N_t-i}=\mathbf{A}_{N_t-i}-\mathbf{B}_{N_t-i}\mathbf{D}^{-1}_{N_t-i}\mathbf{C}_{N_t-i}$ be the \emph{Schur} complement of the matrix $\mathbf{D}_{N_t-i}$ at the $i$th step with the corresponding sub-matrices $\mathbf{A}_{N_t-i}\in\mathbb{C}^{(N_t-i)MN\times (N_t-i)MN}$, $\mathbf{B}_{N_t-i}\in\mathbb{C}^{(N_t-i)MN\times MN}$, $\mathbf{C}_{N_t-i}\in\mathbb{C}^{MN\times (N_t-i)MN}$ and $\mathbf{D}_{N_t-i}\in\mathbb{C}^{MN\times MN}$, where the index $i=1,\ldots,N_t-1$. Since all the matrices $\mathbf{A}_{N_t-i}$, $\mathbf{B}_{N_t-i}$, $\mathbf{C}_{N_t-i}$ and $\mathbf{D}_{N_t-i}$ consist of blocks of $MN\times MN$ diagonal matrices, it follows from \emph{Lemma} \ref{Lemma1} that at the $i$th step, $\mathbf{S}_{N_t-i}\in\mathcal{C}_{(N_t-i)^2,MN}$.  Therefore, $\mathbf{D}_{N_t-i}^{-1}$ always exists and the inverse of $\mathbf{S}_{N_t-i}$ can always be computed by employing the results in \eqref{eq:mimo_otfs_mtx1}. At the final step when $i=N_t-1$, $\mathbf{S}_1=(\mathbf{A}_{1}-\mathbf{B}_{1}\mathbf{D}^{-1}_{1}\mathbf{C}_{1})\in\mathcal{C}_{1,MN}$, and computation of $\mathbf{S}_1^{-1}$ does not need any further partitioning, because $\mathbf{S}_1$ reduces to a diagonal matrix. This completes the proof that inverse of  matrix $\mathbf{X}\in\mathcal{C}_{t^2,MN}$ always exists. The property that the \emph{Schur} complement $\mathbf{S}_{N_t-i}\in\mathcal{C}_{(N_t-i)^2,MN}$ ensures that the inverse $\mathbf{X}^{-1}\in\mathcal{C}_{t^2,MN}$.

\section{}
%\section{Proof of Lemma-\ref{Lemma2_1}}
\label{Lemma2_1_proof}
{Since the matrix $\mathbf{D}\in\mathcal{C}_{N_rN_t,MN}$, computing $\mathbf{D}_{\text{A}}=\mathbf{D}^{H}\mathbf{D}$ for $\text{A}\in\{\text{LZ}\}$, and $\mathbf{D}_{\text{A}}=\mathbf{D}^{H}\mathbf{D}+\rho\mathbf{I}_{N_tMN}$ for $\text{A}\in\{\text{LM}\}$ requires $$[N_t^2N_r+N_t^2(N_r-1)]MN \ \ \text{and }\ [N_t^2N_r+N_t^2(N_r-1)+2N_t]MN$$ operations, respectively.} After that for calculating $\mathbf{D}_{\text{A}}^{-1}$, as shown in Algorithm-\ref{algo:DA_INV}, for $1\leq i\leq N_t-1$, we need to compute  $\mathbf{D}_{N_t-i}^{-1}$ and the corresponding \emph{Schur} complement  $\mathbf{S}_{N_t-i}=\mathbf{A}_{N_t-i}-\mathbf{B}_{N_t-i}\mathbf{D}^{-1}_{N_t-i}\mathbf{C}_{N_t-i}$. This is followed by the computation of $\mathbf{S}_{N_t-i}^{-1}$ using the result in \eqref{eq:mimo_otfs_mtx1} and the backtracking. Independent of the index $i$, $\mathbf{D}_{N_t-i}$ is always an $MN\times MN$ diagonal matrix.  Computing of $\mathbf{D}_{N_t-i}^{-1}$, for each $i$, requires $MN$ multiplications. Since all  matrices $\mathbf{A}_{N_t-i}\in\mathbb{C}^{(N_t-i)MN\times (N_t-i)MN}$, $\mathbf{B}_{N_t-i}\in\mathbb{C}^{(N_t-i)MN\times MN}$, $\mathbf{C}_{N_t-i}\in\mathbb{C}^{MN\times (N_t-i)MN}$  and $\mathbf{D}_{N_t-i}\in\mathbb{C}^{MN\times MN}$ comprise blocks of $MN\times MN$ diagonal matrices, we see that the computation of $\mathbf{S}_{N_t-i}$ costs $(N_t-i)MN+(N_t-i)^2MN+MN$ multiplications and $(N_t-i)^2MN$ additions.  Following the result in \eqref{eq:mimo_otfs_mtx1}, we now need to compute four sub-matrices in terms of the matrices $\mathbf{S}_{N_t-i}^{-1}$, $\mathbf{A}_{N_t-i}$, $\mathbf{B}_{N_t-i}$, $\mathbf{C}_{N_t-i}$ and $\mathbf{D}_{N_t-i}$.  Let $\alpha_{N_t-i}$ be the number of operations required for computing $\mathbf{S}_{N_t-i}^{-1}$. By using the fact from \emph{Lemma} \ref{Lemma2} that $\mathbf{S}_{N_t-i}^{-1}\in\mathcal{C}_{(N_t-i)^2,MN}$, we see that computing $\mathbf{S}_{N_t-i}^{-1}\mathbf{B}_{N_t-i}\mathbf{D}_{N_t-i}^{-1}$ and $\mathbf{D}_{N_t-i}^{-1}\mathbf{C}_{N_t-i}\mathbf{S}_{N_t-i}^{-1}$ require $(N_t-i)^2MN+(N_t-i-1)(N_t-i)MN$ operations each. Computing $\mathbf{D}_{N_t-i}^{-1}+\mathbf{D}_{N_t-i}^{-1}\mathbf{C}_{N_t-i}\mathbf{S}^{-1}_{N_t-i}\mathbf{B}_{N_t-i}\mathbf{D}_{N_t-i}^{-1}$ requires  $(N_t-1)MN+(N_t-i-1)MN+MN$ operations. For each $i$, total operations $\mu_{N_t-i}$ are
 \begin{align}
 \nonumber \mu_{N_t-i} &= (4(N_t-i)^2+3(N_t-i)+2(N_t-i-1)(N_t-i)+(N_t-i-1)+2)MN+\alpha_{N_t-i}\\
 &= (1+2N_t+6N_t^2)MN+(6i^2+12N_ti-2i)MN+\alpha_{N_t-i}.
 \end{align}
We now need to calculate $\alpha_{N_t-i}$ which is the number of operations required for computing the inverse of \emph{Schur} complement $\mathbf{S}_{N_t-i}$.  As $\mathbf{S}^{-1}_{N_t-i}\in\mathcal{C}_{(N_t-i)^2,MN}$, we compute $\mathbf{S}^{-1}_{N_t-i}$, as shown in Algorithm-\ref{algo:DA_INV}, using the result in \eqref{eq:mimo_otfs_mtx1} and backtracking. Thus, $\alpha_{N_t-i}$ can also be calculated by using the procedure explained in the previous paragraph. Consequently, the number of operations required for computing $\mathbf{D}_{\text{A}}^{-1}$ can be evaluated as
\begin{align}
\nonumber \sum_{i=1}^{N_t-1}\mu_{N_t-i} = \sum_{i=1}^{N_t-1}MN\big[(1+2N_t+6N_t^2)+(6i^2+12N_ti-2i)+1\big]=[2N_t^3-2N_t^2+N_t]MN.
\end{align}
Finally, the total number of operations $\mu_{\text{D}_{\text{LZ}}}$ required for computing $\mathbf{D}_{\text{A}}^{-1}$, for $\text{A}\in\text{\{LZ\}}$, can now be evaluated by adding $[N_t^2N_r+N_t^2(N_r-1)]MN$ and $\sum_{i=1}^{N_t-1}\mu_{N_t-i}$, which  yields the desired result for $\mu_{\text{D}_{\text{LZ}}}$ in \emph{Lemma} \ref{Lemma2_1}. Next, the addition of $[N_t^2N_r+N_t^2(N_r-1)+2N_t]MN$ and $\sum_{i=1}^{N_t-1}\mu_{N_t-i}$ gives the desired result for $\mu_{\text{D}_{\text{LM}}}$ in \emph{Lemma} \ref{Lemma2_1}.

\section{}
%\section{Proof of Lemma-\ref{Lemma2_2}}
\label{Lemma2_2_proof}
 Let the matrix $\tilde{\mathbf{D}}_{\text{A}}$ be defined as $\tilde{\mathbf{D}}_{\text{A}}=\mathbf{D}\mathbf{D}_{\text{A}}^{-1}$. Since $\mathbf{D}\in\mathcal{C}_{N_rN_t,MN}$ and, we know from \emph{Lemma} \ref{Lemma2} that $\mathbf{D}_{\text{A}}^{-1}\in\mathcal{C}_{N_t^2,MN}$, calculation of  $\tilde{\mathbf{D}}_{\text{A}}$ requires $[N_t^2N_r+(N_t-1)N_tN_r]MN$ operations. It follows from \emph{Lemma} \ref{Lemma1} that the matrix $\tilde{\mathbf{D}}_{\text{A}}\in\mathcal{C}_{N_rN_t,MN}$.  The receiver matrix $\mathbf{G}_{\text{A}}$ in \eqref{eq:Combiners1} can now be decomposed as $\mathbf{G}_{\text{A}}=\boldsymbol{\Psi}^H_\text{R}\tilde{\mathbf{D}}_{\text{A}}\boldsymbol{\Psi}_\text{T}$. For performing  $\mathbf{G}_{\text{A}}^{H}\mathbf{y}$, we first compute $\tilde{\mathbf{y}}=\boldsymbol{\Psi}_{\text{R}}\mathbf{y}$. Since $\boldsymbol{\Psi}_{\text{R}}$ is a block diagonal matrix whose each block is $\mathbf{F}_M\otimes \mathbf{F}_N$, vector $\tilde{\mathbf{y}}$ is expressed as
\begin{equation}\label{eq:y_tilde}
\tilde{\mathbf{y}}=\left[((\mathbf{F}_M\otimes \mathbf{F}_N)\mathbf{y}_1)^T,\ldots,((\mathbf{F}_M\otimes \mathbf{F}_N)\mathbf{y}_{N_r})^T\right]^T.
\end{equation}
Let $\tilde{\mathbf{Y}}_r$, for $1\leq r\leq N_r$, be the matrices such that $\mbox{vec}(\tilde{\mathbf{Y}}_r)=\tilde{\mathbf{y}}_r$. The vector $(\mathbf{F}_M\otimes \mathbf{F}_N)\mathbf{y}_r$ can then be rewritten as $\mbox{vec}(\mathbf{F}_N\tilde{\mathbf{Y}}_r\mathbf{F}_M^H)$, and can be evaluated by computing $M$-point IDFT along the rows of $\tilde{\mathbf{Y}}_r$ and $N$-point IDFT along the columns of $\tilde{\mathbf{Y}}_r$. Computing $\tilde{\mathbf{y}}$ in \eqref{eq:y_tilde} thus requires $N_r\mathcal{O}(MN\mbox{log}_2MN)$ operations \cite{SurabhiC20}. Computing vector $\mathbf{z}=\tilde{\mathbf{D}}_{\text{A}}^H\tilde{\mathbf{y}}$ requires $N_tN_rMN+N_t(N_r-1)MN$ operations. After this, $\boldsymbol{\Psi}^H_T\mathbf{z}$ can be computed using $N_t\mathcal{O}(MN\mbox{log}_2MN)$ operations. Thus, the number of operations required to process  $\mathbf{y}$~are
\begin{equation}\label{eq:Mu_g}
\mu_{\text{G}_{\text{A}}} = \big[N_t^2N_r+N_tN_r(N_t-1)+N_tN_r+N_t(N_r-1)\big]MN+\big[N_t+N_r\big]\mathcal{O}(MN\mbox{log}_2MN).
\end{equation}
By solving \eqref{eq:Mu_g}, we get the desired result stated in \emph{Lemma} \ref{Lemma2_2}.

\section{}\label{proof_DeltaG}
Using the relation $\widehat{\mathbf{D}}= \mathbf{D}+\Delta\mathbf{D}$, we expand the proposed LM receiver matrix, with channel estimation error, in \eqref{eq:MMSE_COMB_ALT_EST}~as
\begin{align}\label{eq:N_P_D_COV_11}
 \widehat{\mathbf{G}}^{H}_{\text{LM}}&\approx \boldsymbol{\Psi}^{H}_{\text{T}}\big( {\mathbf{D}}^{H} {\mathbf{D}}+\rho\mathbf{I}_{N_{t}MN}+\Delta\mathbf{D}^{H} \mathbf{D}+\mathbf{D}^{H}\Delta\mathbf{D}\big)^{-1} ( \mathbf{D}+\Delta\mathbf{D})^{H}\boldsymbol{\Psi}_{\text{R}}.\\
 &\stackrel{(a)}{=}\boldsymbol{\Psi}^{H}_{\text{T}}( \mathbf{S}^{-1}+ \mathbf{T})^{-1}( \mathbf{D}+\Delta\mathbf{D})^{H}\boldsymbol{\Psi}_{\text{R}}.\label{eq:N_P_D_COV_11_1}
\end{align}
The approximation in \eqref{eq:N_P_D_COV_11} is due to the fact that we neglect the term $\Delta\mathbf{D}^{H}\Delta\mathbf{D}$. This is justified because, as shown in Fig.~\ref{fig:CDF}(b), the probability $P(||\mathbf{\Delta D}^H\mathbf{\Delta D}||_F\ll ||\mathbf{\Delta D}^H\mathbf{D}||_F)$ is close to $1$. This happens because  the  error variance $\sigma^{2}_{e}$ of $\Delta\mathbf{H}$ is small \cite{WangAMMCL07,SinghMJV19}. Equality in $(a)$ follows by substituting $\mathbf{S}^{-1}=\mathbf{D}_{\text{LM}}=\mathbf{D}^H\mathbf{D}+\rho\mathbf{I}_{N_tMN}$ and $\mathbf{T}=\Delta\mathbf{D}^{H} \mathbf{D}+\mathbf{D}^{H}\Delta\mathbf{D}$. {It follows from \cite[Eq. (191)]{petersen2012matrix} that for high SNR the matrix $\mathbf{S}$ can be approximated as $\mathbf{S}\approx \big(\mathbf{D}^H\mathbf{D}\big)^{-1}-\rho \big(\mathbf{D}^H\mathbf{D}\big)^{-2}.$ Each term of the matrix $\mathbf{TS}$  is thus a function of the matrix $\Delta\mathbf{D}$. Let $\lambda_{\text{max}}^{\text{TS}}$ be the maximum eigenvalue of the matrix $\mathbf{TS}$. Fig.~\ref{fig:CDF}(a) shows the empirical complimentary cumulative  distribution function (CCDF) of $|\lambda_{\text{max}}^{\text{TS}}|$. We see that the probability of the random variable $|\lambda_{\text{max}}^{\text{TS}}|\geq 1$ is close to zero. Thus, $(\mathbf{S}^{-1}+\mathbf{T})^{-1}=\mathbf{S}(\mathbf{I}+\mathbf{TS})^{-1}$ can be expanded using Taylor series as \cite{petersen2012matrix}}
\begin{equation}\label{eq:taylor}
( \mathbf{S}^{-1}+ \mathbf{T})^{-1}=\mathbf{S}-\mathbf{S}\mathbf{T}\mathbf{S}+\mathbf{S}(\mathbf{T}\mathbf{S})^{2}-\mathbf{S}(\mathbf{T}\mathbf{S})^{3}+\cdots.
\end{equation}
We see from Fig.~\ref{fig:CDF}(a) that the $P(|\lambda^{\text{TS}}_{\text{max}}|<1)$ is close to 1. Therefore, it follows from \cite{petersen2012matrix} that $ (\mathbf{S}^{-1}+\mathbf{T})^{-1}\approx \mathbf{S}-\mathbf{STS}$. The expression of $\widehat{\mathbf{G}}^{H}_{\text{LM}}$  in \eqref{eq:N_P_D_COV_11_1} can now be simplified~as
\begin{align}\label{eq:N_P_D_COV_2}
\nonumber  \widehat{\mathbf{G}}^{H}_{\text{LM}}&\cong \boldsymbol{\Psi}^{H}_{\text{T}}\mathbf{S}\mathbf{D}^{H}\boldsymbol{\Psi}_{\text{R}}-\boldsymbol{\Psi}^{H}_{\text{T}}\big[\mathbf{S}\big(\Delta\mathbf{D}^{H} \mathbf{D}+\mathbf{D}^{H}\Delta\mathbf{D}\big)\mathbf{S}\mathbf{D}^{H}\\
 &-\mathbf{S}\Delta\mathbf{D}^{H}+\mathbf{S}\big(\Delta\mathbf{D}^{H} \mathbf{D}+\mathbf{D}^{H}\Delta\mathbf{D}\big)\mathbf{S}\Delta\mathbf{D}^{H}\big]\boldsymbol{\Psi}_{\text{R}}.
\end{align}
The first term in \eqref{eq:N_P_D_COV_2} is the proposed LM receiver matrix $\mathbf{G}_{\text{LM}}^{H}$  (see \eqref{eq:Combiners1}). 
We see that in each part of the fourth term, the matrix $\mathbf{\Delta D}$ is multiplied twice. However, in the second and third terms, it is multiplied only once. Since as shown in Fig~\ref{fig:CDF}(b) that $P(||\mathbf{\Delta D}^H\mathbf{\Delta D}||_F\ll ||\mathbf{\Delta D}^H\mathbf{D}||_F)$ is close to $1$, fourth term in \eqref{eq:N_P_D_COV_2} can be ignored when compared with the first, second and third terms.  Using the above observations,  and by comparing \eqref{eq:MMSE_COMB_ALT_EST} and \eqref{eq:N_P_D_COV_2}, we have
$\widehat{\mathbf{G}}^{H}_{\text{LM}}\cong \mathbf{G}_{\text{LM}}^{H}+\Delta\mathbf{G}^{H}_{\text{LM}}$,
where the matrix $\Delta\mathbf{G}_{\text{LM}}$ is given in \eqref{eq:N_P_D_COV_3}.

 \section{}
%\section{Proof of Lemma-\ref{Lemma3}}
\label{Lemma3_proof}
The $(r,t)$th block $\Delta\mathbf{D}_{r,t}\in\mathbb{C}^{MN\times MN}$ of $\Delta\mathbf{D}\in\mathcal{C}_{N_rN_t,MN}$, by using\eqref{eq:Prop2},  can be expressed as
\begin{equation}\label{eq:Lemma3_1}
\Delta\mathbf{D}_{r,t}=\sum_{k=1}^{M}\boldsymbol{\Omega}_{M}^k\otimes\Delta\boldsymbol{\Lambda}^k_{r,t}.
\end{equation}
Here $\Delta\boldsymbol{\Lambda}^k_{r,t}\in\mathbb{C}^{N\times N}$ consists of eigenvalues of the $k$th $N\times N$ circulant block of the sub-matrix $\Delta\mathbf{H}_{r,t}\in\mathbb{C}^{MN\times MN}$ of the error matrix $\Delta\mathbf{H}$. {We know that a circulant matrix can be diagonalized using the DFT matrices. Its eigenvalues can therefore be computed using the DFT of the first row of the circulant matrix \cite{lin2010filter}. Using this property}, the $l$th diagonal entry of  $\Delta\boldsymbol{\Lambda}^k_{r,t}$ is
\begin{equation}\label{eq:Lemma3_2}
\Delta\lambda^k_{r,t,l}=\sum_{n=0}^{N-1}\Delta h_{r,t,n}^k\exp\{j2\pi ln/N\},
\end{equation}
where $\Delta h_{r,t,n}^k$ denotes the $n$th entry of the first row in the $k$th circulant block of the error sub-matrix $\Delta\mathbf{H}_{r,t}$. Let $L_h^k$ be the number of non-zero entries in each row of the $k$th circulant block of the error sub-matrix $\Delta\mathbf{H}_{r,t}$. Since these non-zero entries are i.i.d. with pdf $\mathcal{CN}(0,\sigma^2_e)$, we see that $\Delta\lambda^k_{r,t,l}$ also obeys complex Gaussian distribution with mean $\mathbb{E}[\Delta\lambda^k_{r,t,l}]=0$ and variance  $\mathbb{E}\big[|\Delta\lambda^k_{r,t,l}|^2\big]=\sigma^2_eL_h^k$. In other words, each diagonal entry of $\Delta\boldsymbol{\Lambda}^k_{r,t}$ obeys $\mathcal{CN}(0,\sigma^2_eL_h^k)$. Since $\boldsymbol{\Omega}_{M}^k=\mbox{diag}[1,e^{j2\pi k/M},e^{j4\pi k/M},\ldots,e^{j2\pi k(M-1)/M}]$, it readily follows from \eqref{eq:Lemma3_1} and the properties of matrix Kronecker product that each diagonal element of the eigenvalue matrix $\Delta\mathbf{D}_{r,t}$ also follows the complex Gaussian distribution with mean zero and variance $\sigma^2_d=\sigma^2_e\sum_{k=1}^{M}L_h^k$. The use of the above solutions and the fact that the non-zero entries of error matrix $\Delta\mathbf{H}$ are spatially independent, i.e. independent across the indices $r,t$, yield the desired result in \emph{Lemma} \ref{Lemma3}.

\section{}
%\section{Proof of Lemma-\ref{Lemma4}}
\label{Lemma4_proof}
Let $(i,j)$th block of matrices $\mathbf{X}\in\mathcal{C}_{N_rN_t,MN}$, $\mathbf{Y}\in\mathcal{C}_{N_r^2,MN}$ and $\mathbf{Z}\in\mathcal{C}_{N_t^2,MN}$ be represented as $\mathbf{X}_{i,j}$, $\mathbf{Y}_{i,j}$ and $\mathbf{Z}_{i,j}$, respectively. Since the matrices $\mathbf{X}$ and $\mathbf{Y}$ belong to the sets $\mathcal{C}_{N_rN_t,MN}$ and $\mathcal{C}_{N_r^2,MN}$, respectively, it follows from \emph{Lemma} \ref{Lemma1} that  $\mathbf{X}^H\mathbf{Y}\mathbf{X}$ belongs to the set $\mathcal{C}_{N_t^2,MN}$. The $(k,l)$th block of size $MN\times MN$ of the matrix $\mathbf{X}^H\mathbf{Y}\mathbf{X}$, for $1\leq k,l\leq N_t$, can be computed as 
\begin{equation}
[\mathbf{X}^H\mathbf{Y}\mathbf{X}]_{k,l}=\sum_{j=1}^{N_r}\sum_{i=1}^{N_r}\mathbf{X}_{i,k}^H\mathbf{Y}_{i,j}\mathbf{X}_{j,l}.
\end{equation}
Since each of the sub-matrices $\mathbf{Y}_{i,j}$ and $\mathbf{X}_{j,l}$ is $MN\times MN$ diagonal matrix, we get $\mathbb{E}\big[\mathbf{X}_{i,k}^H\mathbf{Y}_{i,j}\mathbf{X}_{j,l}\big]=\mathbf{Y}_{i,j}\mathbb{E}\big[\mathbf{X}_{i,k}^H\mathbf{X}_{j,l}\big]$. Since all the elements in each block of the matrix $\mathbf{X}$ are zero mean i.i.d with variance $\sigma^2_x$, we get $\mathbb{E}\big[\mathbf{X}_{i,k}^H\mathbf{X}_{j,l}\big]=\sigma^{2}_x \delta_{ij}\delta_{kl}$, where $\delta_{ij}$ denote a Kronecker delta function defined as $\delta_{ij}=1$ if $i=j$ and zero otherwise. Thus, expected value of $[\mathbf{X}^H\mathbf{Y}\mathbf{X}]_{k,l}$ can be evaluated as
\begin{equation}\label{eq:bwe}
\mathbb{E}\big[[\mathbf{X}^H\mathbf{Y}\mathbf{X}]_{k,l}\big]=\sigma^2_x\delta_{k,l}\sum_{i=1}^{N_r}\mathbf{Y}_{i,i}.
\end{equation}
On similar lines, for $1\leq k,l\leq N_r$, 
\begin{equation}\label{eq:bwe1}
\mathbb{E}\big[[\mathbf{X}\mathbf{Z}\mathbf{X}^H]_{(k,l)}\big]=\sigma^2_x\delta_{k,l}\sum_{i=1}^{N_t}\mathbf{Z}_{i,i}.
\end{equation}
By using  \eqref{eq:bwe} and \eqref{eq:bwe1} to evaluate $\mathbb{E}\big[\mathbf{X}^H\mathbf{Y}\mathbf{X}\big]$ and $\mathbb{E}\big[\mathbf{X}\mathbf{Z}\mathbf{X}^H\big]$ respectively, we obtain \eqref{eq:Lemma4_results}.

\section{}\label{Simplify_Rv}

\textit{Simplification of \eqref{eq:N_P_D_COV_4} and \eqref{eq:N_P_D_COV_4_1}:}
{Using \emph{Observation}~\ref{obsrv:5} in \eqref{eq:N_P_D_COV_4} along with \emph{Lemma} \ref{Lemma3}, \emph{Lemma} \ref{Lemma5}, we observe that the third, fifth, seventh and ninth terms in \eqref{eq:N_P_D_COV_4} are zero. By evaluating the first, second, fourth, sixth and eighth terms in \eqref{eq:N_P_D_COV_4} using \emph{Lemma} \ref{Lemma3}, \emph{Lemma} \ref{Lemma4}, the expression for $\mathbb{E}\big[\Delta\mathbf{G}^{H}_{\text{LM}}\mathbf{D}\mathbf{D}^{H}\Delta\mathbf{G}_{\text{LM}}\big]$ can be obtained~as
\begin{align}%\label{eq:N_P_D_COV_5}
\nonumber \mathbb{E}\big[\Delta\mathbf{G}^{H}_{\text{LM}}\mathbf{D}\mathbf{D}^{H}\Delta\mathbf{G}_{\text{LM}}\big] &  =  \sigma^2_d\boldsymbol{\Psi}_{\text{T}}^{H}\Big\{\mathbf{S}(\bar{\mathbf{D}}_{1}- \bar{\mathbf{D}}_{2}+ \bar{\mathbf{D}}_4- \bar{\mathbf{D}}_{6})\mathbf{S}^{H} + \mathbf{D}_3^{H}\bar{\mathbf{D}}_8\mathbf{D}_3 \Big\}\boldsymbol{\Psi}_{\text{T}},
\end{align}
where the block diagonal matrices $\{\bar{\mathbf{D}}_{1},\bar{\mathbf{D}}_{2},\bar{\mathbf{D}}_{4},\bar{\mathbf{D}}_{6}\}\in\mathbb{C}^{N_tMN\times N_tMN}$ and $\bar{\mathbf{D}}_{8}\in\mathbb{C}^{N_rMN\times N_rMN}$ corresponding to the first, second, fourth, sixth and eighth terms of \eqref{eq:N_P_D_COV_4} are computed as $\bar{\mathbf{D}}_{1} = \sum_{i=1}^{N_r}\big(\mathbf{I}_{N_t}\otimes \left[\mathbf{D}_2\mathbf{D}_{1}\mathbf{D}_2^H\right]_{i,i}\big)$, $\bar{\mathbf{D}}_{2} = \sum_{i=1}^{N_r}\big(\mathbf{I}_{N_t}\otimes \left[\mathbf{D}_2\mathbf{D}_{1}\right]_{i,i}\big)$, $\bar{\mathbf{D}}_{4} = \sum_{i=1}^{N_r}\big(\mathbf{I}_{N_t}\otimes \left[\mathbf{D}_{1}\right]_{i,i}\big)$, $\bar{\mathbf{D}}_{6} = \sum_{i=1}^{N_r}\big(\mathbf{I}_{N_t}\otimes \left[\mathbf{D}_{1}\mathbf{D}_{2}^H\right]_{i,i}\big)$ and $\bar{\mathbf{D}}_{8} = \sum_{i=1}^{N_t}\big(\mathbf{I}_{N_r}\otimes \left[\mathbf{D}_{3}^H\mathbf{D}_{1}\mathbf{D}_{3}\right]_{i,i}\big)$, respectively. On the similar lines, the expression of $\mathbb{E}\big[\Delta\mathbf{G}^{H}_{\text{LM}}\Delta\mathbf{G}_{\text{LM}}\big]$ in \eqref{eq:N_P_D_COV_4_1} can be derived  as
\begin{equation}%\label{eq:N_P_D_COV_6}
\nonumber  \mathbb{E}\big[\Delta\mathbf{G}^{H}_{\text{LM}}\Delta\mathbf{G}_{\text{LM}}\big]=\sigma^2_d\boldsymbol{\Psi}_{\text{T}}^{H}\Big\{\mathbf{S}\tilde{\mathbf{D}}_{1}\mathbf{S}^{H}- \mathbf{S}\tilde{\mathbf{D}}_{2}\mathbf{S}^{H}+ N_r\mathbf{S}\mathbf{S}^{H}- \mathbf{S}\tilde{\mathbf{D}}_{6}\mathbf{S}^{H} + \mathbf{D}_3^{H}\tilde{\mathbf{D}}_8\mathbf{D}_3 \Big\}\boldsymbol{\Psi}_{\text{T}}.
\end{equation}
The block diagonal matrices $\{\tilde{\mathbf{D}}_{1},\tilde{\mathbf{D}}_{2},\tilde{\mathbf{D}}_{6}\}\in\mathbb{C}^{N_tMN\times N_tMN}$ and $\tilde{\mathbf{D}}_{8}\in\mathbb{C}^{N_rMN\times N_rMN}$, corresponding to the first, second, sixth and eighth terms of \eqref{eq:N_P_D_COV_4_1}, are given as $\tilde{\mathbf{D}}_{1}=\sum_{i=1}^{N_r}\big(\mathbf{I}_{N_t}\otimes \left[\mathbf{D}_2\mathbf{D}_2^H\right]_{i,i}\big)$, $\tilde{\mathbf{D}}_{2}=\sum_{i=1}^{N_r}\big(\mathbf{I}_{N_t}\otimes \left[\mathbf{D}_2\right]_{i,i}\big)$, $\tilde{\mathbf{D}}_{6}=\sum_{i=1}^{N_r}\big(\mathbf{I}_{N_t}\otimes \left[\mathbf{D}_2^H\right]_{i,i}\big)$ and $\tilde{\mathbf{D}}_{8}=\sum_{i=1}^{N_t}\big(\mathbf{I}_{N_r}\otimes \left[\mathbf{D}_3^H\mathbf{D}_3\right]_{i,i}\big)$.}

\ifCLASSOPTIONcaptionsoff
  \newpage
\fi

%\vspace{-1.cm}

\bibliographystyle{IEEEtran}
\bibliography{IEEEabrv,OTFS_ZF_MMSE}

\begin{thebibliography}{10}
\providecommand{\url}[1]{#1}
\csname url@samestyle\endcsname
\providecommand{\newblock}{\relax}
\providecommand{\bibinfo}[2]{#2}
\providecommand{\BIBentrySTDinterwordspacing}{\spaceskip=0pt\relax}
\providecommand{\BIBentryALTinterwordstretchfactor}{4}
\providecommand{\BIBentryALTinterwordspacing}{\spaceskip=\fontdimen2\font plus
\BIBentryALTinterwordstretchfactor\fontdimen3\font minus
  \fontdimen4\font\relax}
\providecommand{\BIBforeignlanguage}[2]{{%
\expandafter\ifx\csname l@#1\endcsname\relax
\typeout{** WARNING: IEEEtran.bst: No hyphenation pattern has been}%
\typeout{** loaded for the language `#1'. Using the pattern for}%
\typeout{** the default language instead.}%
\else
\language=\csname l@#1\endcsname
\fi
#2}}
\providecommand{\BIBdecl}{\relax}
\BIBdecl

\bibitem{DBLP:books/cu/G2005}
A.~Goldsmith, \emph{Wireless Communications}.\hskip 1em plus 0.5em minus
  0.4em\relax Cambridge University Press, 2005.

\bibitem{WangPMZ06}
T.~Wang, J.~G. Proakis, E.~Masry, and J.~R. Zeidler, ``Performance degradation
  of {OFDM} systems due to {Doppler} spreading,'' \emph{{IEEE} Trans. Wireless
  Commun.}, vol.~5, no.~6, pp. 1422--1432, Jun 2006.

\bibitem{DBLP:journals/corr/abs-1802-02623}
\BIBentryALTinterwordspacing
R.~Hadani and A.~Monk, ``{OTFS:} {A} new generation of modulation addressing
  the challenges of {5G},'' \emph{CoRR}, vol. abs/1802.02623, 2018. [Online].
  Available: \url{http://arxiv.org/abs/1802.02623}
\BIBentrySTDinterwordspacing

\bibitem{HadaniRTMGMC17}
R.~Hadani, S.~Rakib, M.~Tsatsanis, A.~Monk, A.~J. Goldsmith, A.~F. Molisch, and
  A.~R. Calderbank, ``Orthogonal time frequency space modulation,'' in
  \emph{{IEEE}, {WCNC}, San Francisco, CA, USA, March 19-22}, 2017, pp. 1--6.

\bibitem{DBLP:books/ieee/Jakes74}
W.~C. Jakes, \emph{Microwave Mobile Communications}.\hskip 1em plus 0.5em minus
  0.4em\relax Wiley/IEEE Press, 1974.

\bibitem{RavitejaPH19}
P.~Raviteja, K.~T. Phan, and Y.~Hong, ``Embedded pilot-aided channel estimation
  for {OTFS} in delay-{Doppler} channels,'' \emph{{IEEE} Trans. Vehicular
  Technol.}, vol.~68, no.~5, pp. 4906--4917, Mar. 2019.

\bibitem{SurabhiC20}
G.~D. Surabhi and A.~Chockalingam, ``Low-complexity linear equalization for
  {OTFS} modulation,'' \emph{{IEEE} Commun. Lett.}, vol.~24, no.~2, pp.
  330--334, Feb. 2020.

\bibitem{SurabhiAC19}
G.~D. Surabhi, R.~M. Augustine, and A.~Chockalingam, ``On the diversity of
  uncoded {OTFS} modulation in doubly-dispersive channels,'' \emph{{IEEE}
  Trans. Wireless Commun.}, vol.~18, no.~6, pp. 3049--3063, Apr. 2019.

\bibitem{raviteja2018interference}
P.~Raviteja, K.~T. Phan, Y.~Hong, and E.~Viterbo, ``Interference cancellation
  and iterative detection for orthogonal time frequency space modulation,''
  \emph{{IEEE} Trans. Wireless Commun.}, vol.~17, no.~10, pp. 6501--6515, Aug.
  2018.

\bibitem{RavitejaPJHV18}
P.~Raviteja, K.~T. Phan, Q.~Jin, Y.~Hong, and E.~Viterbo, ``Low-complexity
  iterative detection for orthogonal time frequency space modulation,'' in
  \emph{{IEEE}, {WCNC}, 2018, Barcelona, Spain, April 15-18}, 2018, pp. 1--6.

\bibitem{RamachandranC18}
M.~K. Ramachandran and A.~Chockalingam, ``{MIMO-OTFS} in high-{Doppler} fading
  channels: Signal detection and channel estimation,'' in \emph{{IEEE}
  {GLOBECOM}, Abu Dhabi, United Arab Emirates, December 9-13}, 2018, pp.
  206--212.

\bibitem{SurabhiRC19}
G.~D. Surabhi, M.~K. Ramachandran, and A.~Chockalingam, ``{OTFS} modulation
  with phase noise in mmwave communications,'' in \emph{89th {IEEE} {VTC}
  Spring, Kuala Lumpur, Malaysia, April 28 - May 1}, 2019, pp. 1--5.

\bibitem{MuraliC18}
K.~R. Murali and A.~Chockalingam, ``On {OTFS} modulation for high-doppler
  fading channels,'' in \emph{2018 Information Theory and Applications
  Workshop, {ITA} 2018, San Diego, CA, USA, February 11-16, 2018}, 2018, pp.
  1--10.

\bibitem{TiwariDR19}
S.~Tiwari, S.~S. Das, and V.~Rangamgari, ``Low complexity {LMMSE} receiver for
  {OTFS},'' \emph{{IEEE} Commun. Lett.}, vol.~23, no.~12, pp. 2205--2209, Oct.
  2019.

\bibitem{cheng2019low}
J.~Cheng, H.~Gao, W.~Xu, Z.~Bie, and Y.~Lu, ``Low-complexity linear equalizers
  for {OTFS} exploiting two-dimensional fast {Fourier} transform,'' \emph{arXiv
  preprint arXiv:1909.00524}, 2019.

\bibitem{surabhi2020low}
G.~Surabhi and A.~Chockalingam, ``Low-complexity linear equalization for
  2$\times$ 2 {MIMO-OTFS} signals,'' in \emph{IEEE 21st International Workshop
  on Signal Processing Advances in Wireless Communications (SPAWC)}, 2020, pp.
  1--5.

\bibitem{chockalingam2014large}
A.~Chockalingam and B.~S. Rajan, \emph{Large {MIMO} systems}.\hskip 1em plus
  0.5em minus 0.4em\relax Cambridge University Press, 2014.

\bibitem{golub2012matrix}
G.~H. Golub and C.~F. Van~Loan, \emph{Matrix computations}.\hskip 1em plus
  0.5em minus 0.4em\relax JHU press, 2012, vol.~3.

\bibitem{FishGHSS13}
A.~Fish, S.~Gurevich, R.~Hadani, A.~M. Sayeed, and O.~Schwartz,
  ``Delay-{Doppler} channel estimation in almost linear complexity,''
  \emph{{IEEE} Trans. Inf. Theory}, vol.~59, no.~11, pp. 7632--7644, Jul. 2013.

\bibitem{DingSFP19a}
Z.~Ding, R.~Schober, P.~Fan, and H.~V. Poor, ``{OTFS-NOMA:} an efficient
  approach for exploiting heterogenous user mobility profiles,'' \emph{{IEEE}
  Trans. Commun.}, vol.~67, no.~11, pp. 7950--7965, Aug. 2019.

\bibitem{SinghMJVH20}
P.~Singh, H.~B. Mishra, A.~K. Jagannatham, K.~Vasudevan, and L.~Hanzo, ``Uplink
  sum-rate and power scaling laws for multi-user massive {MIMO-FBMC} systems,''
  \emph{{IEEE} Trans. Commun.}, vol.~68, no.~1, pp. 161--176, Jan. 2020.

\bibitem{kra2012circulant}
I.~Kra and S.~R. Simanca, ``On circulant matrices,'' \emph{Notices of the AMS},
  vol.~59, no.~3, pp. 368--377, 2012.

\bibitem{RottenbergMHL19}
F.~Rottenberg, X.~Mestre, F.~Horlin, and J.~Louveaux, ``Efficient equalization
  of time-varying channels in {MIMO} {OFDM} systems,'' \emph{{IEEE} Trans.
  Signal Process.}, vol.~67, no.~21, pp. 5583--5595, 2019.

\bibitem{petersen2012matrix}
K.~B. Petersen and M.~S. Pedersen, ``The matrix cookbook, nov 2012,'' \emph{URL
  http://www2. imm. dtu. dk/pubdb/p. php}, vol. 3274, 2012.

\bibitem{lu2002inverses}
T.-T. Lu and S.-H. Shiou, ``Inverses of 2$\times$ 2 block matrices,''
  \emph{Computers \& Mathematics with Applications}, vol.~43, no. 1-2, pp.
  119--129, 2002.

\bibitem{WangAMMCL07}
C.~Wang, E.~K.~S. Au, R.~D. Murch, W.~H. Mow, R.~S. Cheng, and V.~K.~N. Lau,
  ``On the performance of the {MIMO} zero-forcing receiver in the presence of
  channel estimation error,'' \emph{{IEEE} Trans. Wireless Commun.}, vol.~6,
  no.~3, pp. 805--810, Mar. 2007.

\bibitem{SinghMJV19}
P.~Singh, H.~B. Mishra, A.~K. Jagannatham, and K.~Vasudevan, ``Semi-blind,
  training, and data-aided channel estimation schemes for {MIMO-FBMC-OQAM}
  systems,'' \emph{{IEEE} Trans. Signal Process.}, vol.~67, no.~18, pp.
  4668--4682, Jul. 2019.

\bibitem{HassibiH03}
B.~Hassibi and B.~M. Hochwald, ``How much training is needed in
  multiple-antenna wireless links?'' \emph{{IEEE} Trans. Inf. Theory}, vol.~49,
  no.~4, pp. 951--963, Apr. 2003.

\bibitem{lin2010filter}
Y.-P. Lin, S.-M. Phoong, and P.~Vaidyanathan, \emph{Filter bank transceivers
  for {OFDM} and {DMT} systems}.\hskip 1em plus 0.5em minus 0.4em\relax
  Cambridge University Press, 2010.

\end{thebibliography}
\end{document}